\documentclass[11pt, reqno]{amsart}
\pdfoutput=1
\usepackage{amssymb, amsmath, euscript, verbatim, array, url, mathrsfs, amscd, color, slashed, tabularx} 
\usepackage[all]{xy}

\newtheorem{Theorem}{Theorem}[section]
\newtheorem{Lemma}[Theorem]{Lemma}

\theoremstyle{definition}
\newtheorem{Definition}[Theorem]{Definition}

\newtheorem{Remark}[Theorem]{Remark}
\newcommand{\Proof}{\textbf{Proof}\hspace{0.3cm}}
\newcommand{\End}{\ensuremath{\;\square}\\}

\numberwithin{equation}{section}

\newcommand{\mr}[1]{\mathrm{#1}}
\newcommand{\mc}[1]{\mathcal{#1}}

\newcommand{\R}{\mathbb{R}}

\newcommand{\A}{\mathcal{A}}

\newcommand{\vp}{\varphi}

\newcommand{\pd}{\partial}

\renewcommand{\L}{\mathcal{L}}

\newcommand{\eps}{\epsilon}

\renewcommand{\S}{\mathcal{S}}

\renewcommand{\Re}{\mathrm{Re}\,}

\renewcommand{\S}{\mathcal{S}}

\newcommand{\g}{\frak{g}}

\renewcommand{\S}{S}
\newcommand{\bW}{\mathbb{W}}
\renewcommand{\L}{\mathcal{L}}

\begin{document}

  \title[Path Integrals]{The Perturbative Approach to Path Integrals: \\A Succinct Mathematical Treatment}
  \author{Timothy Nguyen}
    \address{Michign State University \\ 619 Red Cedar Road \\ East Lansing, MI 48824}
\date{\today}
\email{timothyn@math.msu.edu}

\begin{abstract}
We study finite-dimensional integrals in a way that elucidates the mathematical meaning behind the formal manipulations of path integrals occurring in quantum field theory. This involves a proper understanding of how Wick's theorem allows one to evaluate integrals perturbatively, i.e., as a series expansion in a formal parameter irrespective of convergence properties. We establish invariance properties of such a Wick expansion under coordinate changes and the action of a Lie group of symmetries, and we use this to study essential features of path integral manipulations, including coordinate changes, Ward identities, Schwinger-Dyson equations, Faddeev-Popov gauge-fixing, and eliminating fields by their equation of motion. We also discuss the asymptotic nature of the Wick expansion and the implications this has for defining path integrals perturbatively and nonperturbatively.
\end{abstract}

\maketitle

\tableofcontents

\section*{Introduction}

Quantum field theory often makes use of manipulations of path integrals that are without a proper mathematical definition and hence have only a formal meaning. These computations are usually presented in a manner that makes it difficult to ascertain the mathematical nature of these operations, other than that they are inspired from familiar properties of finite-dimensional integrals. The purpose of this paper is to explain in a large class of important examples how formal path integral manipulations constitute \textit{notation} for otherwise mathematically well-defined procedures. While some practitioners of quantum field theory are aware of these issues (see e.g. \cite{ZJ}), considerations of rigor are highly non-uniform in the quantum field theoretic literature, making it a challenge to navigate for both beginner and expert alike. The present note arose from the need to have a simple explanation of the mathematics behind perturbative quantum field theory, one that is both rigorous and sufficiently comprehensive to cover examples of interest.

A basic prototype for the kinds of integrals one studies in quantum mechanics and quantum field theory is the one-dimensional integral
\begin{equation}
I(\lambda) = \int dx\, e^{-\frac{1}{2}x^2 - \lambda x^n}
\label{eq:I-intro}\end{equation}
with $\lambda$ a real parameter and $n \geq 3$. The standard procedure of expanding $e^{-\lambda x^n}$ as a Taylor series and then integrating term by term against the Gaussian measure $dx e^{-\frac{1}{2}x^2}$ yields a well-defined formal power series in $\lambda$ irrespective of whether $I(\lambda)$ converges (which holds only for $\lambda \geq 0$ when $n$ is even and $\lambda=0$ when $n$ is odd). In particular, we can formally manipulate the integrand occurring in $I(\lambda)$, such as performing a change of variables, and still produce a corresponding formal series in $\lambda$ using the same method. 

This procedure of evaluating an integral as a formal power series in the relevant coupling constants is the guiding principle behind perturbative quantum field theory. In this paper, we refer to such a procedure as the \textit{Wick expansion}. It arises from a formal application of the saddle point approximation and does not require the underlying integral to converge. In the infinite-dimensional setting of quantum field theory, the Wick expansion appropriately generalized (which leads to the familiar regularization and renormalization of Feynman diagrams) provides a \textit{definition} of path integrals. Such a definition bypasses the difficulties (or even impossibility) of constructing the appropriate measures on infinite-dimensional spaces needed to evaluate a path integral as an honest integral. Nevertheless, one refers to the path integral as being defined \textit{perturbatively} through this formal series method, as though the series were known to approximate path integrals as honest integrals. (This is inspired from the fact that for $n$ even, the Wick  expansion for $I(\lambda)$ yields an asymptotic series for $I(\lambda)$, regarded as an honest function of $\lambda$, as $\lambda \rightarrow 0^+$.) As a consequence, the many results in quantum field theory obtained from formal manipulations of path integrals naturally arouse suspicion. However, based on the finite-dimensional case above, the way out of this confusion is to have a clear separation between integration, which is analytic, and the Wick expansion, which is algebraic.

This paper provides a mathematical study of the Wick expansion of finite-dimensional integrals in a way that provides insight into (perturbative) path integrals in quantum field theory. Namely, we establish invariance properties of Wick expansions with respect to standard calculus manipulations of the integrals used to define them and show how this allows one to make mathematical meaning of formal path integral manipulations in quantum field theory. It is helpful to illustrate this approach via the following schema:\pagebreak

$$\;$$
\begin{center}
Finite-Dimensional Integrals
\end{center}
$$
 \xymatrix@M=.1in@R=.75in{ I(\lambda) \ar@{->}^-{\textrm{Wick expansion}}[rrr] \ar@{-->}_-{\textrm{integration}\,}[d] & & & \fbox{\parbox{1in}{\begin{center}formal series \\ in $\lambda$\end{center}}} \\
 \fbox{\parbox{1.75in}{\begin{center}function of $\lambda$\end{center}}} \ar@{-->}_-{\hspace{.5in}\textrm{asymptotic series}}[urrr] & &
}
$$


\vspace{.4in}

\begin{center}
 Perturbative Path Integrals
\end{center}
$$
 \xymatrix@M=.1in@R=.75in{ Z(\lambda) = \int D\phi\, e^{\int \left(-\frac{1}{2}|\nabla \phi|^2 + \lambda \phi^n\right)} \ar@{->}^-{\textrm{Wick expansion}}[rrr]  & & & \fbox{\parbox{1in}{\begin{center}formal series \\ in $\lambda$\end{center}}}
 }
$$
\vspace{.1in}
\begin{center}
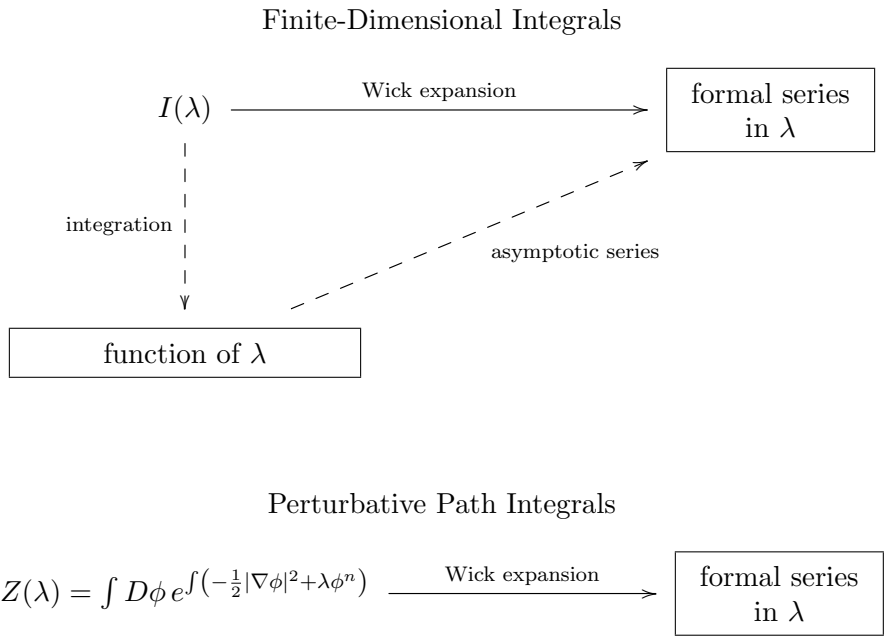
 Figure 1. The Wick expansion and its relationship to integrals.  \vspace{.4in}
\end{center}


\noindent Thus, while integral manipulations can be analyzed in the usual way when integration is well-defined, they can also (and always) be analyzed using the Wick expansion.

This paper is organized as follows. In Section 1, we define the Wick expansion and show that it is independent of the choice of coordinates used to define it. The latter is important for showing that the Wick expansion can be defined on manifolds. It should be emphasized that the Wick expansion is an expansion about an arbitrary nondegenerate critical point (the Hessian can be indefinite), so that such an expansion is not always a Gaussian approximation. We also study the more general Morse-Bott case in which we Wick expand about critical submanifolds. In Section 2, we discuss the Wick expansion in the presence of symmetries. Here, a suitable gauge-fixing procedure is needed to define the Wick expansion. In doing so, we obtain a rigorous formulation of the Faddeev-Popov gauge-fixing procedure in the finite-dimensional setting, which works whenever we have a volume preserving action of a (not necessarily compact) Lie group on a manifold. 
In fact, we emphasize the there are two types of Faddeev-Popov procedures (we call them the \textit{slice} and \textit{weighted} versions), and we clarify both of these in the context of integration and the Wick expansion in ways that are crucial for understanding quantum gauge theories. In Section 3, we relate the Wick expansion to the asymptotics of integrals (this is the diagonal arrow in Figure 1), thereby relating purely formal algebraic manipulations to analytic properties of integrals which are convergent and satisfy additional hypotheses. This provides the justification for why the Wick expansion is a sensible algebraic object to associate to an integral. Finally in Section 4 we discuss how to interpret our results in the context of quantum field theory. Namely, we discuss the Wick expansion in light of the regularization procedures typically used in quantum field theory -- either from regulating the propagator, the dimension, or using a lattice. From this, we can suitably interpret a variety of formal manipulations of path integrals defined perturbatively in a mathematically rigorous manner. By comparison, standard textbook treatments of these procedures (see e.g. \cite[Ch 16.4]{PS} \cite[Ch 15.7]{Wein}) interpret them via integration, which is often illegitimate (both literally in the infinite-dimensional setting and by analogy with the finite-dimensional setting). This is particularly true of gauge-fixing procedures, for which the blur between integration and the Wick expansion hampers a rigorous understanding as our discussion will show. We believe the examples we discuss are only a few among many that achieve clarity (and perhaps even conceptual correction) through the methodology presented in this paper, which is centered around the Wick expansion. We conclude with a brief discussion concerning perturbative versus constructive quantum field theory.

Altogether, our treatment separates in a simple and elegant manner the analytic difficulties of working in infinite dimensions from the distinction between integration and the Wick expansion already occurring in finite dimensions. In other words, we clearly distinguish between analysis (integration) and algebra (Wick expansion). Typically, analytic approaches to quantum field theory (e.g. those arising from Wightman and Osterwalder-Schrader axioms \cite{GJ}) do not resemble the treatment of perturbative quantum field theory that appears in textbooks \cite{Col, PS, Wein}, whose methods have proven to be remarkably successful in real world applications. On the other hand, algebraic approaches to quantum field theory are often limited to or are only concerned with producing formal series expansions, which even in finite dimensions, do not capture integration without further analytical considerations. We hope our explicit delineation of algebraic and analytic methods provides clarity to the limitations that various approaches to quantum field theory have. Perhaps a helpful analogy would be to regard algebraic methods that produce a formal perturbative series as providing a ``weak'' construction of a path integral; analytic methods which attempt to boost such a formal series to a well-defined function then provide a ``strong'' construction. As is true in many other contexts, it is conceptually sound to have a clear distinction between weak and strong constructions of an object.

In order to keep this note succint, we wrote it as a compromise between comprehensiveness and brevity with emphasis on simplicity. For alternative treatments to some of the issues raised here, see e.g. \cite{JF, Sch}.

\section{The Wick Expansion}

Consider the integral
\begin{align}
  I = \int d^dx\,f(x)e^{-S(x)/\hbar} \label{I}
\end{align}
on $\R^d$ where $S$ and $f$ are complex-valued functions. We always assume that our functions are smooth, i.e., infinitely differentiable. For the situations relevant to physics, we regard $S(x)$ as an action, $f(x)$ an observable, and the parameter $\hbar$ a real or complex parameter. Moreover, we think of $\hbar$ as being small, since $\hbar \to 0$ is to be regarded as a semiclassical limit of (\ref{I}). A rescaling of (\ref{I}) shows that $\hbar$ can also be regarded as a perturbative parameter, since for example, letting $S(x) = \frac{1}{2}x^2 + x^n$, $f(x) \equiv 1$, and $x \mapsto \hbar^{1/2}x$, we recover the integral considered in the introduction with $\lambda = \hbar^{n/2-1}$.

Under the appropriate hypotheses, the integral $I$ is convergent and becomes a function of $\hbar$. On the other hand, one can treat $\hbar$ as a purely formal parameter, in which case $I$ is a formal object consisting of an integral sign and an integrand. One can then attempt to define various series expansions in $\hbar$ from $I$ using purely algebraic rules. One such expansion is the Wick expansion, which takes as an additional source of input a nondegenerate critical point $x_0$ of $S(x)$. Under appropriate additional assumptions (see Section 3), the Wick expansion about $x_0$ provides the asymptotics of $I$ when $f(x)$ is a bump function localized around a nondegenerate critical point of $S(x)$. However, the Wick expansion is always a well-defined series regardless of the convergence properties of $I$, since it depends only on the derivatives of the integrand of $I$ at $x_0$.

In fact, the significance of the algebraic nature of the Wick expansion goes beyond just bypassing convergence issues of $I$, since the latter can always be made convergent by making $f(x)$ compactly supported. In the infinite dimensional setting of quantum field theory, an honest integral $I$ is replaced with a formal path integral. One often does not know how to make sense of the integrand of such a path integral, since it may lack a well-defined construction as an honest measure. However, the Wick expansion, suitably interpreted, provides an algebraic way to integrate formally such an ill-defined measure. This is based on the analogy between the Wick expansion and integration that will become apparent in what follows. We return to the bearing these observations have on quantum field theory in Section 4.

We now proceed to define the Wick expansion. Consider a nondegenerate critical point $x_0$ of $S(x)$, i.e., one for which the Hessian of $S(x)$ at $x_0$ is nondegenerate. Explicitly, we can write
\begin{align}
  S(x) = S(x_0) + \frac{1}{2}A(x - x_0, x-x_0) + O(|x-x_0|^3) \label{critpt}
\end{align}
for $x$ near $x_0$, where the symmetric bilinear pairing $A = A(\cdot,\cdot)$ is the Hessian. Thus, we can write
\begin{align}
  e^{-S(x)/\hbar} = e^{-S(x_0)/\hbar}e^{-A(x - x_0, x-x_0)/2\hbar} \cdot e^{\bar S(x)/\hbar},\label{Sdecomp}
\end{align}
the product of a Gaussian and an ``interaction" term. (For the time being, we assume $A$ is positive definite, though we will remove this assumption shortly.) In (\ref{I}), we can expand $e^{\bar S(x)/\hbar}f(x)$ as a Taylor series centered at $x_0$ and then integrate term by term against the Gaussian measure $d^dx\, e^{-S(x_0)/\hbar}e^{-A(x - x_0, x-x_0))/2\hbar}$. Aside from an overall normalization factor, we obtain a formal power series in $\hbar$. Indeed, one can easily see this from the rescaling $x \mapsto x_0 + \hbar^{1/2}(x-x_0)$. This makes the Gaussian measure proportional to $\hbar^{d/2}$, and then the integration of polynomials of even degree only picks up even powers of $\hbar^{1/2}$.

The above construction yields the Wick expansion of $I$ about $x_0$ in case $A$ is positive definite. However, it can be generalized to $A$ nondegenerate as follows. The first step is to recall a result that goes by the name of Wick's Theorem, which provides a convenient, combinatorial formula for evaluating integrals of polynomials against Gaussian measures. The easiest way to describe a polynomial function on a vector space is to pick a basis. On $\R^d$, we can pick a basis $e_1, \ldots, e_d$ and a corresponding dual basis $x^1, \cdots, x^d$ of coordinate monomials. We abbreviate
$$d^dx = dx^1\cdots dx^d$$
for the corresponding density, which we assume to be the same density that appears in (\ref{I}). We can describe a normalized Gaussian measure in terms of the matrix
$$A_{ij} = A(e_i,e_j)$$
via
$$d\mu_A = dx^1\cdots dx^d\left(\frac{\det A_{ij}}{(2\pi)^{d}}\right)^{1/2}e^{-A(x,x)/2}.$$
Let $A^{ij}$ denote the inverse matrix of $A_{ij}$.

\begin{Theorem}\label{ThmWick}
  (Wick's Theorem) For $A$ positive definite, we have
  \begin{equation}
    \int d\mu_A \,x^{i_1}\cdots x^{i_{2m}} = \frac{1}{2^{m}m!}\sum_{\sigma \in S_{2m}}A^{i_{\sigma(1)}i_{\sigma(2)}}\cdots A^{i_{\sigma(2m-1)}i_{\sigma(2m)}}. \label{wick}
  \end{equation}
\end{Theorem}

This formula can be encoded pictorially through the use of Feynman diagrams, for which the contractions of the $A^{ij}$ into the slots of $x^{i_1}\cdots x^{i_{2m}}$ are encoded through incidence relations among edges and vertices of graphs. Further details can be found e.g. in \cite{Cos, PS} or other textbooks on quantum field theory.

Note that while the left-hand side of (\ref{wick}) is analytic in nature, the right-hand side is purely combinatorial. In particular, one can define the right-hand side for any nonsingular (complex-valued) matrix $A$. One can also regard the right-hand side of (\ref{wick}) as providing an analytic continuation of the left-hand side to the space of nonsingular matrices. Hence, we make the following definition:

\begin{Definition}
  Given a nondegenerate, symmetric bilinear pairing $A$, define the \textit{Wick operator} $\bW_{A}$ to be the linear functional on the space of polynomials given by the following formula:
  \begin{align}\bW_A(P) =
  \begin{cases}\displaystyle
    \frac{1}{2^{m}m!}\sum_{\sigma \in S_{2m}}A^{i_{\sigma(1)}i_{\sigma(2)}}\cdots A^{i_{\sigma(2m-1)}i_{\sigma(2m)}} & P = x^{i_1}\cdots x^{i_{2m}}\\
    0 & P \textrm{ is odd.}
  \end{cases}\label{WO}
  \end{align}
\end{Definition}

Although the above formula makes use of a basis, it is easy to see that the definition of $\bW_A$ depends only on the bilinear pairing $A$. The terms $A^{ij}$ appearing in the Wick formula (\ref{WO}) are called \textit{Wick contractions}.

Wick's Theorem tells us that the integration of polynomials against the Gaussian measure $d\mu_A$ coincides with the Wick operator $\bW_A$ for $A$ positive definite. Thus, the series expansion which we described above, which involved integrating the Taylor series of $f(x)e^{\bar S(x)/\hbar}$ against the Gaussian $d^dx e^{-S(x_0)/\hbar}e^{-A(x - x_0, x-x_0)/2\hbar}$, can be defined using the Wick operator $\bW_A$ instead of integration. But since the Wick operator is well-defined for any nondegenerate $A$, this allows us to extend the definition of such an expansion to the case when $S(x)$ has a critical point with arbitrary nondegenerate Hessian:

\begin{Definition}\label{DefWE}
  Consider the integral $I$ along with a choice of local coordinates near a nondegenerate critical point $x_0$ of $S(x)$. This yields a fixed decomposition of the integrand of $I$ into the product of the coordinate density $d^dx$ and the function $f(x)e^{-S(x)/\hbar}$ near $x_0$. Represent $e^{-S(x)/\hbar}$ as in (\ref{Sdecomp}) and Taylor expand
  \begin{align}
    f(x_0 + \hbar^{1/2}(x-x_0))e^{\bar S(x_0 + \hbar^{1/2}(x-x_0))/\hbar} \label{int}
  \end{align}
  about $x = x_0$ to obtain a power series $\sum_{k =0}^\infty \hbar^{k/2} P_{k/2}(x-x_0)$
  grouped by powers of $\hbar$. Then the \textit{Wick expansion} of $I$ about $x_0$ is the formal series in $\hbar$ given by
  \begin{equation}
    W_{x_0}(\hbar) = \left(\frac{(2\pi \hbar)^d}{\det A_{ij}}\right)^{1/2}e^{-S(x_0)/\hbar}\sum_{k = 0}^\infty c_k \hbar^k.
  \end{equation}
  where $c_k = \bW_{A}(P_k)$.
\end{Definition}

A priori, the Wick expansion depends on the choice of coordinates used to define it. Indeed, a different choice of coordinates leads to a Jacobian factor and a different sequence of polynomials occurring in the Taylor expansion of (\ref{int}). Nevertheless, Theorem \ref{LemmaCoord} tells us that the Wick expansion yields a series independent of the coordinate system used to construct it. First, we prove the following lemma:

\begin{Lemma}\label{LemmaIBP}
  The Wick expansion of a total derivative is zero.
\end{Lemma}

\Proof We need to show that the Wick expansion of
\begin{align}
  I &= \int d^dx \,\pd_{x^i}[f(x)e^{-S(x)/\hbar}] \nonumber\\
    &= \hbar^{-1}\int d^dx\, e^{-S(x)/\hbar}[\hbar \pd_{x^i}f(x) -\pd_{x^i} S(x)f(x)] \label{eq:deriv}
\end{align}
vanishes. Here, the Wick expansion is obtained by applying Definition \ref{DefWE} to the final expression above (the $\hbar^{-1}$ is just an overall constant factor). Using the splitting (\ref{Sdecomp}), it follows that in the case of $A$ positive definite, the vanishing of the Wick expansion follows from
\begin{equation}
\int d^dx\, \pd_{x^i} \left[P(x)e^{-A(x,x)/2}\right] = 0
\end{equation}
for arbitrary polynomials $P(x)$. For general nondegenerate $A$, one has to establish the algebraic analogue of the above equation, namely
\begin{align}
  \bW_A\left(\pd_{x^i}P(x) -A_{ij}x_j P(x)\right) = 0. \label{Wvanisih}
\end{align}
Verifying this identity is straightforward.\End

\begin{Theorem}\label{LemmaCoord}
  The Wick expansion about a nondegenerate critical point $x_0$ is independent of the choice of coordinates.
\end{Theorem}

\Proof Let $\Phi: \R^d \to \R^d$ be a diffeomorphism. We want to show that the Wick expansion of $d^dx\,f(x)e^{-\S(x)/\hbar}$ and $\Phi^*(d^dx\,f(x)e^{-\S(x)/\hbar})$ about $x_0$ and $\Phi^{-1}(x_0)$, respectively, are equal as series expansions in $\hbar$. Without loss of generality, we can suppose $x_0 = 0$ and $\Phi(0) = 0$. Moreover, it is easy to see that the Wick expansion is invariant under a linear change of coordinates, since linear maps preserve polynomial degree and hence the Wick formula (\ref{WO}). Thus, we may further suppose that $D\Phi$ at $x=0$ is the identity. It follows that the $1$-parameter family of maps
$$\Phi_t(x) = (1 - t)x + t\Phi(x)$$
is a local diffeomorphism in a neighborhood of $0$ for all $t \in [0,1]$. It suffices to show that the Wick expansion of $\frac{d}{dt}\Phi_t^*(d^dx\,f(x)e^{-\S(x)/\hbar})$ is identically zero for all $t$. Letting
\begin{align}
  V_t(x) = \frac{d}{ds}\bigg|_{s=t}\Phi_s(\Phi_t^{-1}(x)) \label{Vt}
\end{align}
denote the time-dependent vector field associated to the flow $\Phi_t$, then
\begin{align}
  \frac{d}{dt}\Phi_t^*(d^dx\,f(x)e^{-\S(x)/\hbar}) &= \Phi_t^*\left(\mathcal{L}_{V_t} (d^dx\,f(x)e^{-\S(x)/\hbar})\right) \nonumber \\
  &= \Phi_t^*\left(d\iota_{V_t}(d^dx\,f(x)e^{-\S(x)/\hbar})\right) \nonumber \\
  &= d \iota_{\Phi_t^*(V_t)} \Phi_t^*\left(d^dx\,f(x)e^{-\S(x)/\hbar}\right), \label{cartan}
\end{align}
where $\mathcal{L}_{V_t}$ is the Lie derivative, and in the second line we used the Cartan formula $\mathcal{L}_{V_t} = d \iota_{V_t}$ for the Lie derivative of a differential form of top degree. Since $\Phi_t(x) = x + O(|x|^2)$ for all $t$, then the $\iota_{\Phi_t^*(V_t)}\Phi_t^*\left(d^dx\,f(x)e^{-\S(x)/\hbar}\right)$ are all of the form $d^dx\,f_t(x)e^{-A(x,x)/2\hbar}e^{-\bar S_t(x)/\hbar}$ for some $t$-dependent functions $\bar S_t$ and $f_t$. Hence, the integrands (\ref{cartan}) are all integrands for which we may Wick expand. By Lemma \ref{LemmaIBP}, these Wick expansions are all zero.\End

\begin{Remark}\label{RemFormal}
  The Wick expansion only depends on the derivatives of the integrand of $I$ at $x_0$, i.e., the infinite jet. Thus, one should really work in the category of formal power series about $x_0$ (e.g. changes of coordinates need only be invertible formal power series). We will leave this understanding implicit and instead maintain more geometric terminology throughout the paper by considering all objects as smooth.
\end{Remark}

\subsection{The Morse-Bott case}\label{SecMB}

The coordinate invariance of the Wick expansion allows us to generalize our definition of it on Euclidean space to the setting of smooth manifolds. Consider the integral
\begin{equation}
  I = \int_M dV f(x)e^{-S(x)/\hbar} \label{IM}
\end{equation}
where $dV$ is a differential form of top degree\footnote{Our discussion generalizes straightforwardly to (and ought to be phrased in terms of) densities, but to keep our discussion a bit simpler and more familiar, we restrict ourselves to working with differential forms on orientable manifolds.} on $M$. Note that the presence of $f(x)$ is redundant since it can be grouped with $dV$, but we keep $f(x)$ separate since one often keeps some reference $dV$ fixed while varying $f(x)$. Moreover, the presence of the $\hbar$ parameter in the exponential means that the ($\hbar$-independent) function $S(x)$ is uniquely-defined,  and we can refer to critical sets for $S(x)$ as critical sets of (the integrand of) $I$. Recall that a submanifold $Z \subset M$ is \textit{critical} for $S$ if $(d\S)_x = 0$ for all $x \in Z$.

When $Z = x_0$ is a nondegenerate critical point, we can work in local coordinates near $x_0$ and Wick expand about $x_0$. This yields a well-defined series in $\hbar$ independent of the coordinate system chosen by Theorem \ref{LemmaCoord}. We now want to consider the case when $Z$ is \textit{Morse-Bott nondegenerate}, i.e., for every $x \in Z$, the kernel of the Hessian of $S$ at $x$ equals $T_xZ$. It is important to make this generalization since the types of integrals one often encounters are those for which $S(x)$ has some moduli of stationary configurations.

Given any Morse-Bott nondegenerate critical submanifold $Z \subset M$, consider a small tubular neighborhood $\tilde Z$ of $Z$. By regarding $\tilde Z$ as a fibration over $Z$ by disks (say by picking a Riemannian metric on $M$ and using the exponential map in the direction orthogonal to $Z$), then the corresponding bundle projection $\pi: \tilde Z \to Z$ allows us to perform fiber integration \cite{BT}.  This is a map $\pi_*$ sending top-degree differential forms on $\tilde Z$ to top-degree differential forms on $Z$, which is essentially integration along the fiber directions of $\tilde Z$. Moreover, fiber integration is volume-preserving:
\begin{align}
  \int_{\tilde Z} d\mu = \int_Z \pi_*(d\mu). \label{eq:fint}
\end{align}

Given the analogy between the Wick expansion and integration, we define a Wick expansion version of fiber integration, which we call Wick fiber integration. As one might expect, Wick fiber integration maps top-degree forms on $\tilde Z$ to top-degree forms on $Z$ valued in a formal series in $\hbar$. This operation depends only on the local behavior near $Z$ and hence extends to a map on top-degree forms on $M$.

We define Wick fiber integration as follows. Given the bundle projection $\pi$, we can choose local coordinates adapted to $\pi$ in the following sense. Given $p \in Z$, we can choose coordinates $(x,y) = (x^1,\ldots, x^r, y^1,\ldots, y^{d-r})$ on an open set $U$ containing $p$ such that $Z \cap U = \{y^1 = \ldots = y^{d-r} = 0\}$ and $\pi: \tilde Z \cap U \to Z \cap U$ is projection onto the $x$-coordinates. We refer to $x$ and $y$ as horizontal and vertical directions, respectively. The Morse-Bott nondegeneracy condition on $Z$ implies that in these coordinates
\begin{align}
  \S(x,y) = S(x,0) + \frac{1}{2}A_x(y,y) + \bar S_x(y),
\end{align}
where $A_x(y, y)$ is a nondegenerate $x$-dependent quadratic form of the variables $y^1, \ldots, y^{d-r}$ and $\bar S_x(y) = O(y^3)$. Thus, with $p \in Z$ fixed, we can Wick expand $(\ref{IM})$ about $p$ with respect to the fiber variables $y^1,\ldots, y^{d-r}$, thereby obtaining a series in $\hbar$ valued in a top-degree differential form on $Z$. Indeed, the Wick expansion eliminates the $y$-variables and leaves the $x$-variables remaining.

Abbreviate the integrand of (\ref{IM}) by
\begin{align}
  \mc{V} = dVf(x)e^{-\S(x)/\hbar}. \label{V}
\end{align}

\begin{Definition}\label{Def:WFI}
  Let $Z \subset M$ be a submanifold and choose a fiber bundle structure $\pi: \tilde Z \to Z$. \textit{Wick fiber integration} $\pi^W_*$ is the map which sends $\mc{V}$, a top-degree differential form (\ref{V}) with $Z$ a Morse-Bott nondegenerate critical set, to $\pi^W_*(\mc{V})$, a top-degree differential form on $Z$ valued in a formal series in $\hbar$, by performing fiberwise Wick expansion of $\mc{V}$ along each point of $Z$.
\end{Definition}

As before, we want to show that our definition of Wick fiber integration is independent of the choice of coordinates used to perform the Wick expansion along individual fibers.

\begin{Lemma}
  The Wick fiber integration map $\pi^W_*$ is well-defined, i.e., it is independent of the choice of coordinates used to define it.
\end{Lemma}

\Proof Given coordinates $(x,y)$ adapted to $\pi$, another coordinate system adapted to $\pi$ is obtained via a local diffeomorphism $(\Phi^h(x), \Phi^v(x,y))$, where $\Phi^h$ and $\Phi^v$ are the horizontal and vertical components of $\Phi$. Note that $\Phi^h$ does not depend on $y$, since $\Phi$ preserves the fibers of $\pi$. The Wick expansion only depends on the Taylor series of all objects involved in $y$-variables. Thus, using Theorem \ref{LemmaCoord}, we can suppose $\Phi^v(x,y)$ is the identity map on the $y$-variables up to translation, i.e. $\Phi^v(x,y) = \Phi^v(x) + y$. In other words, $\Phi$ maps the horizontal slice $Z = \{(x,0)\}$ to the set of points $\{(\Phi^h(x),\Phi^v(x))\}$ and it vertically translates fibers accordingly. Any such translation diffeomorphism commutes with the Wick expansion in the $y$-variables, i.e., one can Wick expand then push-forward by $\Phi$ or else pushforward by $\Phi$ and then Wick expand. This shows that $\pi_*^W$ is well-defined, since it transforms appropriately under diffeomorphisms.\End

Another way of reformulating the above is the following:

\begin{Lemma}\label{ThmWickCov}
  Wick fiber integration is covariant. That is, if $\Phi: M \to M$ is a diffeomorphism intertwining fiber bundle structures $\pi$ and $\Phi_*\pi := \Phi \circ \pi \circ \Phi^{-1}$ for $Z$ and $\Phi(Z)$, respectively, then
  \begin{align}
    \Phi_*(\pi^W_*(\mc{V})) = (\Phi_*\pi)^W_*(\Phi_*(\mc{V})). \label{cov}
  \end{align}
\end{Lemma}

Fiber integration maps top-degree differential forms on a total space to top-degree differential forms on the base such that the resulting map on cohomology is independent of the homotopy class of the fiber bundle structure. This is because if $\pi_t = \pi \circ \Phi_t^*$, where $\Phi_t$ is a one-parameter family of diffeomorphisms, then letting $V_t$ denote the vector field associated to the flow $\Phi_t$ as in (\ref{Vt}), we have
\begin{align}
  \frac{d}{dt}(\pi_t)_* &= \pi_* \circ \frac{d}{dt}\Phi_t^* \nonumber \\
  &= \pi_* \circ \Phi_t^* \circ d \iota_{V_t} \nonumber \\
  &= d \circ \pi_* \circ \Phi_t^* \circ \iota_{V_t}. \label{dpi}
\end{align}
Here, we used that fiber integration commutes with the exterior derivative.

Wick fiber integration, being a formal analogue of fiber integration, satisfies the following property:

\begin{Lemma}\label{LemmaWickHomotope}
  The terms of the Wick fiber integration $\pi_*^W(\mc{V})$ have well-defined cohomology classes that are independent of the choice of fiber bundle structure $\pi$.
\end{Lemma}

\Proof Given a smooth family $\pi_t$ of bundle structures, then by repeating the derivation (\ref{dpi}) with Wick fiber integration in place of fiber integration, we have that $\frac{d}{dt}(\pi_t)_*^W(\mc{V})$ at $p \in Z$ consists of the Wick expansion along $\pi_t^{-1}(p)$ of a total derivative. Near $Z$, $d$ splits into a (time-dependent) fiber component and a component $d_Z$ tangential to $Z$. Thus $\frac{d}{dt}(\pi_t)_*^W(\mc{V})$ is equal to a total $d_Z$ derivative, since the fiber component of $d$ is annihilated (by Lemma \ref{LemmaIBP}) and $d_Z$ commutes with Wick expansion in the fiber directions. This establishes the theorem for all fiber bundle structures homotopic to $\pi$. Now we recall Remark \ref{RemFormal} and note that Wick fiber integration with respect to $\pi$ only depends on the Taylor expansion of $\pi$ in the directions transverse to $Z$ (in particular, with respect to the fiber directions with respect to some fixed bundle structure); in the Morse-Bott situation, this Taylor expansion has coefficients that are functions on $Z$. It is thus enough to show that the fiberwise Taylor expansions along $Z$ of any two fiber bundle structures can be homotoped.

To see this, note that given a fixed $\pi: \tilde Z \to Z$, we can always regard $\pi$ as a vector bundle projection $\pi: NZ \to Z$ (where $NZ$ is the normal bundle to $Z$ with respect to some arbitrary metric on $M$) restricted to $\tilde Z \subset NZ$. With this normalization, the Taylor series of $\pi$ along the fibers of $NZ$ is trivial. A general fiber bundle map $\pi': \tilde Z \to Z$ has a nontrivial Taylor series along the fibers of $NZ$, since the latter fibers will not coincide with the fibers of $\pi'$. The derivatives along $Z$ of $\pi'$ in the fiber directions of $NZ$ are sections of the bundle $\prod_{k \geq 1}\mr{End}(\mr{Sym}^k(NZ), TZ)$. This is a contractible space, with the zero section corresponding to $\pi$. It follows that we can homotope the fiberwise Taylor expansions of any two fiber bundle structures.\End

As formula (\ref{eq:fint}) shows, fiber integration along $Z$ splits integration in a neighborhood of $Z$ into a fiber direction and a remaining direction parallel to $Z$. Our definition of the Wick expansion in the Morse-Bott setting is defined accordingly. Namely, the Wick expansion about $Z$ involves a Wick fiber integration in the fiber directions, and then a residual integration along $Z$. More precisely, we have the following:

\begin{Definition}\label{DefMB}
  Suppose $Z$ is a compact Morse-Bott nondegenerate critical set for $\mc{V}$. Define the \textit{Wick expansion} of $I$ about $Z$ to be the formal series in $\hbar$ defined by
  \begin{align}
    W_Z(\hbar) = \int_Z \pi^W_*\Big(dVf(x)e^{-S(x)/\hbar}\Big),
  \end{align}
  where $\pi$ is any fiber bundle structure for $Z$. Lemma \ref{LemmaWickHomotope} implies this definition is independent of the choice of $\pi$.
\end{Definition}
We emphasize that the integration over $Z$ in the above is an honest integration, in which case we must make some assumptions to ensure that such an integration is well-defined. One cannot perform a residual Wick expansion on $Z$, since $\pi^W_*\Big(dVf(x)e^{-S(x)/\hbar}\Big)$ is of the form $e^{C/\hbar}$ times a power series in $\hbar$, with $C$ the constant value of $S$ along $Z$. Thus, we make the assumption that $Z$ is compact.


\begin{Theorem}
  The Wick expansion $W_Z(\hbar)$ of $I$ depends only on the diffeomorphism class of the integrand of $I$.
\end{Theorem}

\Proof This is an automatic consequence of Lemma \ref{ThmWickCov}.\End

\noindent Note that because of the above theorem, it was justifiable to speak of the Wick expansion as being a function of $I$ instead of its integrand, since in case $I$ is convergent, its numerical value depends only on the diffeomorphism class of its integrand as well.

Altogether, we see the formal analogies between the Wick expansion and integration. Both are coordinate independent and satisfy analogous properties with regards to integration by parts and fiber integration. We can summarize these observations in the following informal principle:\\


\noindent \textbf{Formal Integration Principle: }\textit{Any natural identity on integrals yield a corresponding identity on Wick expansions.}\\

The next section presents a further manifestation of this principle.

\section{The Wick Expansion and Gauge-Fixing}\label{Sec:WEGF}

We now consider integrands that are invariant under a (not necessarily compact) Lie group $G$ of symmetries. Such a symmetry is regarded as a redundancy among the variables occuring in the integrand.  In the context of integration, this means we should factor out the contribution of $G$ to the associated integral. In the context of the Wick expansion, this means we need to eliminate the degeneracies in the action arising from $G$-invariance. For both these situations, the Faddeev-Popov procedure allows us to perform the required operations just described in terms of a choice of gauge-fixing, which we define in a moment. This procedure actually has two different incarnations which we refer to as the \textit{slice} and \textit{weighted} Faddeev-Popov procedures. The distinction between these two methods and whether they should be applied in the context of integration or the Wick expansion is blurred in the literature, and our goal here is to give a clear and unified treatment that emphasizes their different features. This will pay great dividends later when we study the Faddeev-Popov procedure in the setting of quantum field theory. 

We first work in the context of integration. Moreover, we begin by describing the slice Faddeev-Popov method. The setup is as follows. 
We have a manifold $M$ equipped with a volume form $dV$ that is preserved under a left-action of $G$. We suppose $G$ acts freely on $M$ else we can restrict to a subgroup of $G$. Consider the integral
$$I = \int_M dVf(x)$$
where $f$ is a $G$-invariant function. To eliminate the redundancy arising from $G$-invariance, we can choose a local slice $\mc{S}$ for the $G$-action, that is, a locally closed submanifold $\mc{S} \subset M$ which is transverse to the action of $G$ (i.e.\! the $G$-orbit $G\cdot \mc{S}$ through $\mc{S}$ is diffeomorphic to $G \times \mc{S}$).  We refer to this procedure of choosing a local slice as \textit{gauge-fixing}, since its analog in infinite dimensions is what one does when choosing a gauge-fixing condition.

Having chosen a local slice $\mc{S}$, we can replace the part of the integral $I$ over $G\cdot \mc{S}$ with a suitably weighted integral over $\mc{S}$.  The most notable feature of this gauge-fixing procedure is the presence of a determinant. Such a determinant can be described implicitly using merely the fact that $\mc{S}$ is transverse to the $G$-action or else explicitly with the aid of a $G$-invariant Riemannian metric on $M$ such that $dV$ is the associated Riemannian volume form. This determinant takes into account that the volumes of $G$-orbits vary in $M$, so that a weight is needed   
when passing from an integral over $M$ to that on $\mc{S}$. (For the case of $G$ noncompact, one instead considers the ``ratio" of volumes of different $G$-orbits). Such a determinant is referred to as the \textit{Faddeev-Popov determinant}. 

We first prove an ``implicit'' version of the slice Faddeev-Popov method and then derive an explicit Riemannian version. While the implicit version makes no auxiliary choices, it is the explicit version that is more useful in practice for doing computations. In what follows, we assume without loss of generality that $M$ is globally a product
\begin{equation}
 M \cong G \times \mc{S}.\label{eq:globalslice}
\end{equation}
This is always true locally, and we can assume it is true globally in the context of integration by using a $G$-invariant partition of unity. We refer to $\mc{S}$ in (\ref{eq:globalslice}) as being a global slice.

Given a global slice $\mc{S}$, the multiplication map
\begin{align*}
  \rho: G \times \mc{S} & \to M \\
  (g,w) & \mapsto g\cdot w,
\end{align*}
is a $G$-equivariant diffeomorphism, where $G$ acts on $G \times \mc{S}$ by left multiplication on the first factor. Pick any left-invariant volume form $dV_G$ on $G$. In doing so, we can define the volume form $dV_{\mc{S}}^{M/G}$ on $\mc{S}$ via
$$\rho^*(dV) = dV_G \times dV_{\mc{S}}^{M/G}.$$
Indeed, because $\rho$ yields an identification $T_{g\cdot w}M = T_gG \oplus T_w\mc{S}$ at every point $g\cdot w \in M$, an element in the top exterior power of $T_{g\cdot w}^*M$ and $T_g^*G$ determines one for $T_w^*{\mc{S}}$.

\begin{Theorem}\label{ThmFP.i}
 (slice Faddeev-Popov formula, implicit version)
For any two global slices $\mc{S}$ and $\mc{S}'$, we have
\begin{align}
  \int_{\mc{S}}dV_{\mc{S}}^{M/G}f(w) = \int_{\mc{S}'}dV_{\mc{S}'}^{M/G}f(w). \label{eq:divideG}
\end{align}
Furthermore, if $G$ is compact, then
  \begin{equation}
    \int_M dV f(x) = \mr{Vol}(G)\int_{\mc{S}}dV_{\mc{S}}^{M/G}f(w). \label{FP.i}
  \end{equation}
\end{Theorem}

\Proof The case $G$ compact follows readily from the fact that we can integrate along the fibers of the projection $\pi: M \to \mc{S}$ and $dV_{\mc{S}}^{M/G}$ satisfies
$$\pi_*(dV) = \mr{Vol}(G)dV_{\mc{S}}^{M/G}.$$
Thus, (\ref{eq:divideG}) and (\ref{FP.i}) follows from (\ref{eq:fint}). 

For general $G$, we proceed as follows. Suppose we have two global slices $\mc{S}$ and $\mc{S}'$. This means there exists a ``gauge transformation" $\theta: \mc{S} \to G$ that yields for us an induced diffeomorphism
\begin{align}
  \Theta: \mc{S} &\to \mc{S}' \nonumber \\
w & \mapsto \theta(w)\cdot w \label{Theta}
\end{align}
relating the slices via the $G$-action. We want to show that
\begin{align}
  \Theta^*(dV_{\mc{S'}}^{M/G}) = dV_{\mc{S}}^{M/G}. \label{Theta-pres}
\end{align}
Define the map
\begin{align*}
L_\theta: M & \to M \\
g\cdot w & \mapsto g\cdot w' = g\theta(w)\cdot w.
\end{align*}
We have the commutative diagram
$$\begin{CD}
  G \times \mc{S} @>\mr{id} \times {\Theta}>> G \times \mc{S}' \\
  @V{\rho}VV  @VV{\rho}V \\
  M @>>L_\theta > M
\end{CD}$$
The vertical maps are volume-preserving by definition, since the volume forms on the product spaces are induced by pullback from multiplication. Thus, showing (\ref{Theta-pres}) is equivalent to showing that $L_\theta$ is volume preserving. 

Given $x \in M$, we have the map $\rho_x: G \to M$, $g \mapsto g\cdot x$ determining the $G$-orbit through $x$. Define its differential to be
\begin{align*}
\iota_x: \g &\to M\\
\iota_x(X) &= \frac{d}{dt}\bigg|_{t=0}e^{tX}\cdot x, \qquad X \in \g. 
\end{align*}
giving via the infinitesimal action of $G$ on $M$. In this way, each tangent space to the $G$-orbit through $x$ can be identified with the space of right-invariant vector fields of $G$ via the map $\iota$.

We can write an arbitrary tangent vector $u \in T_{g\cdot w}M$ in terms of its components with respect to the $G$-invariant distributions given by the tangent spaces to the $G$-orbit $G\cdot w$ and $\mc{S}$,
$$u = g_*\iota_w(X) + g_*(v), \qquad X \in \g,\; v \in T_w\mc{S}.$$
Likewise, we can write a tangent vector at $T_{g\theta(w)\cdot w}M$ as $(g\theta(w))_*\iota_w(X) + g_*\big((D_w\Theta)(v)\big)$. We have $(D_w\Theta)(v) = R_w(v) + \theta(w)_*(v)$, were $R_w$ maps $v$ into the image of $\theta(w)_*\circ \iota_w$. Piecing all these decompositions together, the derivative of $L_\theta(g\cdot w) = g\cdot\Theta(w)$ at $g\cdot w$ is given by
$$D_{g\cdot w}L_\theta\Big(g_*\iota_w(X), g_*(v)\Big) = \begin{pmatrix}
  (g\theta(w)g^{-1})_*(g_*\iota_w(\cdot)) & g_*R_w(\cdot) \\
  0 & (g\theta(w)g^{-1})_*(g_*(\cdot))
\end{pmatrix}\begin{pmatrix}
  X \\ v
\end{pmatrix}.$$
Being upper triangular, it determines the same map on top-degree forms as $(g\theta(w)g^{-1})_*$. The latter is volume-preserving since left multiplication by $G$ is volume-preserving. Thus, $L_\theta$ is volume preserving, thereby establishing (\ref{Theta-pres}).

Hence, for $f$ a $G$-invariant function,
$$\int_{\mc{S}} dV_{\mc{S}}^{M/G}f(w) =
\int_{\mc{S}}\Theta^*\Big(dV_{\mc{S}'}^{M/G}f(w')\Big) = \int_{\mc{S'}}dV_{\mc{S}'}^{M/G}f(w').\End$$

The disadvantage with the above formulation of the Faddeev-Popov procedure is that it does not express the volume form $dV_{\mc{S}}^{M/G}$ in a very explicit manner (especially for generalization to quantum field theory). However, if we endow $M$ with a $G$-invariant metric, with $dV$ the associated $G$-invariant Riemannian volume, we can describe $dV_{\mc{S}}^{M/G}$ explicitly as follows. The Riemannian metric on $M$ restricts to a Riemannian metric on $\mc{S}$ and so induces its own volume form $dV_{\mc{S}}$ on $\mc{S}$.  

\begin{Definition}
  The \textit{Faddeev-Popov determinant} $J_{\mc{S}}(w)$ is the function on $\mc{S}$ defined by
  \begin{equation}
   dV^{M/G}_{\mc{S},w} = J_{\mc{S}}(w)dV_{\mc{S},w}, \qquad w \in \mc{S}. 
  \end{equation}
\end{Definition}

An explicit characterization of the Faddeev-Popov determinant, which is indicative of its name, is as follows. The $G$-invariant volume form $dV_G$ on $G$ is determined by the volume form $dV_\g$ it induces on its Lie algebra. 
For $w \in \mc{S}$, we can compose $\iota_w$ with the orthogonal projection onto $T_w^\bot\mc{S}$, the orthogonal complement of $T_w\mc{S}$ inside $T_wM$. This yields for us the map $\iota_w^\bot: \g \to T_w^\bot\mc{S}$. It is an isomorphism since $\mc{S}$ is transverse to the $G$-action. Let $dV_{\mc{S}}^\bot$ and $dV_{\mc{S}}$ denote the volume forms on $T^\bot_w\mc{S}$ and $\mc{S}$, respectively, determined by the metric induced from $M$. Then we have
\begin{equation}
    (\iota_w^\bot)^*(dV_{\mc{S},w}^\bot) = J_{\mc{S}}(w)dV_\g, \qquad w \in \mc{S}. \label{defFP}
  \end{equation} 
Explicitly, let $e_j^*$ be a basis for $\g^*$ such that $dV_\g = \wedge e_j^*$ and let $v_i^*(x)$ be an orthonormal coframe for $\mc{S}$, i.e., the $v_i^*(w)$ form an orthonormal basis of $\mr{Ann}(T_w\mc{S}) \subset T_x^*M$. In particular, $dV_{\mc{S},w}^\bot = \wedge v_i^*$. Define $A_{ij}(w)$ by
$$(\iota_w^\bot)^*(v_i^*) = A_{ij}(w)e_j^*.$$
Then
\begin{equation}
  J_{\mc{S}}(w) = \det A_{ij}(w). \label{detA}
\end{equation}
  
An automatic consequence of Theorem \ref{ThmFP.i} and the definition of $J_{\mc{S}}(w)$ is the following Riemanian version:

\begin{Theorem} \label{ThmFP.R} (slice Faddeev-Popov formula, Riemannian version)
For any two global slices $\mc{S}$ and $\mc{S}'$, we have
\begin{align}
  \int_{\mc{S}}dV_{\mc{S}}J_{\mc{S}}(w)f(w) = \int_{\mc{S}'}dV_{\mc{S}'}J_{\mc{S}'}(w)f(w). \label{eq:divideG.R}
\end{align}
Furthermore, if $G$ is compact, then
  \begin{equation}
    \int_M dV f(x) = \mr{Vol}(G)\int_{\mc{S}}dV_{\mc{S}}J_{\mc{S}}(w)f(w). \label{FP}
  \end{equation}
\end{Theorem}

\begin{Remark}
  In some formulations of the Faddeev-Popov formula, a square root of a determinant appears instead of a determinant. Namely, one has the formula
  \begin{equation}
    J_{\mc{S}}(w) = {\det}^{1/2}((\iota_w^\bot)^*\iota_w^\bot) \label{det1/2}
  \end{equation}
  where $(\iota_w^\bot)^*$ is the adjoint of $\iota_w^\bot: \g \to T_w^\bot\mc{S}$. The square root occurs because the volume form associated to a metric tensor $g_{ij}$ receives a factor of $\det^{1/2}(g_{ij})$, and the pullback by $\iota_w^\bot$  of the metric on $T_w^\bot\mc{S}$ is given by
  $$\left<\iota_w^\bot u, \iota_w^\bot v\right>_{T_w^\bot\mc{S}} = \left<u, (\iota_w^\bot)^*\iota_w^\bot v\right>_\g, \qquad u,v \in \g.$$
\end{Remark}
\vspace{4ex}

Next, we discuss the weighted Faddeev-Popov procedure which is a generalization of the slice method. We discuss both these methods since there is a practical distinction between these two methods when applied to quantum field theory, as we discuss in Section \ref{SecGF}. For the weighted method, instead of \textit{fixing} the gauge by having an integral over $\mc{S}$, we \textit{break} gauge-invariance by introducing a suitable weight function on $M$ which is non-constant along gauge-orbits. The slice method is recovered by letting this weight function tend to the delta-current determined by the chosen slice.

More precisely, let $F: M \to \mc{Q}$ be a gauge-fixing function from $M$ to some target manifold $\mc{Q}$ such that 
\begin{enumerate}
 \item $\mc{S} = F^{-1}(q_0)$ with $q_0$ a regular value (i.e. $dF_w: T_wM \to T_{q_0}\mc{Q}$ is surjective for all $w \in \mc{S}$);
 \item $F$ restricted to some (and hence every) gauge-orbit has nonzero degree.
\end{enumerate}
Here, the degree $\mr{deg}(F)$, in the case of $G$ and $\mc{Q}$ compact, is defined in the usual way: if
\begin{align}
F \circ \rho_w : G &\to \mc{Q}\nonumber \\
g &\mapsto F(g\cdot w) \qquad w \in \mc{S} \label{eq:Forbit}
\end{align}
pulls back a unit-volume top-degree form on $\mathcal{Q}$ to $d$ times such a form on $G$, then $\mr{deg}(F) = d$ is the degree. The independence of the choice of $w$ follows from the fact that the degree is a homotopy invariant. One can also adapt the above setup to the situation of noncompact $G$ by requiring (\ref{eq:Forbit}) to be proper and working with compactly supported top-degree forms, but we will not do so here since we are ultimately interested in the Wick version for which $G$ can be arbitrary.

Pick a unit-volume top-degree form $dq$ on $\mc{Q}$ and a weight function $\vp: \mc{Q} \to \R$ such that $\int dq\,\vp(q) = 1$. The following tells us that by inserting into $I$ the weight $\vp (F(x))$ along with the determinant $\det (dF \circ \iota_x)$ defined via 
$$(dF \circ \iota_x)^*(dq_{F(x)}) = \det (dF \circ \iota_x)dV_\g,$$
we get a weighted integral independent of the choice of $F$ and $\vp$ up to an overall constant:
 
\begin{Theorem} \label{ThmFPw} (weighted Faddeev-Popov formula)
 We have
 \begin{equation}
\int_M dV \vp(F(x))\det (dF \circ \iota_x) f(x) = \mr{deg}(F)\int_{\mc{S}}dV^{M/G}_{\mc{S}}f(w) \label{eq:FPw}  
 \end{equation}
\end{Theorem}

\Proof Pull back the integral over $M$ in (\ref{eq:FPw}) to $\mc{S} \times M$ via $\rho: G \times \mc{S} \to M$. Fixing $w \in \mc{S}$, when we integrate over its $G$ orbit, we have to perform the integral
\begin{align*}
\int_G dV_G\, \vp(F \circ \rho_w)\det (dF \circ \iota_w) &= \int_G (F \circ \rho_w)^*(dq\,\vp(q))\\
&= \deg(F) \int_{\mc{Q}} dq\,\vp(q)\\
&= \deg(F). 
\end{align*}
Peforming the remaining integral over $\mc{S}$ yields the result.\End

\begin{Remark}
  Let us take a closer look at how our formulation of the Faddeev-Popov formula relates to the usual informal expression involving delta functions, which is expressed as
  \begin{equation}
    \int_{M} dV\delta(F(x))\det(dF \circ \iota_x)f(x). \label{FPdelta}
  \end{equation} 
  By slight abuse of terminology, one also refers to $\det(dF \circ \iota_x)$ as the Faddeev-Popov determinant, although the term $\delta(F(x))$, which restricts the integral to $\mc{S}$, contributes a factor coming from the differential of $F$. We can regard (\ref{FPdelta}) as arising from (\ref{eq:FPw}) by letting $\vp$ become a delta function at $q_0$. We can also compare (\ref{FPdelta}) directly with (\ref{eq:divideG.R}) to see how to make a direct translation between these two formulations. Here, for concreteness, we suppose $F: M \to \g$, with $\mc{S} = F^{-1}(0)$. In local coordinates, write the $1$-form $dF^i$ restricted to $\mc{S}$ in terms of an orthonormal conormal frame $e_j^*$ for $\mc{S}$, i.e. $dF^i = B^i_j e_j^*$ for some matrix $B^i_j$. Then $\delta(F(x))$  acquires an inverse determinant factor $(\det B^i_j)^{-1}$, which cancels the corresponding factor of $\det B^i_j$ arising in $\det(dF \circ \iota_x)$. The overall determinant factor we obtain is (\ref{detA}). In this way, we see that transformation properties of the $\delta(F(x))$ and $\det(dF \circ \iota_x)$ conspire to make (\ref{FPdelta}) independent of the choice $F$ used to define the slice $\mc{S}$.
\end{Remark}

We now intertwine the Faddeev-Popov formula with the Wick expansion. Consider the integral
\begin{equation}
I = \int_M dVf(x)e^{-S(x)/\hbar} \label{eq:Isec2}
\end{equation}
where $S$ and $f$ are both $G$-invariant. We want to obtain a Wick expansion of this integral about a critical subset of $S$. Such a Wick expansion requires a choice of gauge-fixing: we need to eliminate degenerate directions arising from the $G$-invariance of $S(x)$ in order to obtain a splitting of $S(x)$ into a nondegenerate quadratic part and a higher order interaction part.

Note that every critical set $Z$ of $S(x)$ is $G$-invariant. We say that $Z$ is \textit{Morse-Bott $G$-nondegenerate} if on the quotient $M/G$, the set $Z/G$ is \textit{Morse-Bott nondegenerate} with respect to the induced function $S(x)$. Given such a function $S(x)$, we can define a Wick expansion of (\ref{eq:Isec2}) by adapting the Faddeev-Popov procedure defined for integration. We can adapt either the slice or the weighted procedures, but unlike in the case of integration, in which a degree term appears that distinguishes these methods, the Wick expansion version of these procedures yields the same output. 

Let us begin with the slice version. On a small $G$-invariant patch of $M$, instead of $I$ we consider the integral
$$I_{\mc{S}} = \int_{\mc{S}} dV_{\mc{S}}^{M/G}f(w)e^{-S(w)/\hbar}.$$
It is this integral that we can Wick expand using the methods of Section \ref{SecMB}. Indeed,
$$Z_{\mc{S}} := Z \cap \mc{S}$$
is a Morse-Bott nondegenerate level set for $S$ restricted to $\mc{S}$ and we can choose a Morse-Bott bundle structure $\pi: \tilde Z_{\mc{S}} \to Z_S$ as before, where $\tilde Z_{\mc{S}}$ is a tubular neighborhood of $Z_{\mc{S}}$ inside $\mc{S}$. Given any two gauge-fixing slices $\mc{S}$ and $\mc{S'}$, we can always relate them via a unique map $\Theta: \mc{S} \to \mc{S'}$ determined by the $G$-action as in (\ref{Theta}).
The next theorem tells us that Wick fiber integration along a gauge-fixed critical set $Z_{\mc{S}}$ is covariant with respect to changes of gauge.

\begin{Theorem}\label{ThmOnshell}
Let $S$ and $f$ be $G$-invariant functions and $Z$ a Morse-Bott $G$-nondenerate submanifold. Pick a slice $\mc{S}$ and a fiber bundle structure $\pi$ for $Z_{\mc{S}}$. Let $\mc{S'}$ be any other slice and let $\pi'$ be the fiber bundle structure for $Z_{\mc{S'}}$ obtained from $\pi$ by the $G$-action relating $\mc{S}$ to $\mc{S}'$. In other words, we have the commutative diagram
$$\begin{CD}
  \tilde Z_{\mc{S}} @>{\Theta}>> \tilde Z_{\mc{S}'} \\
  @V{\pi}VV  @VV{\pi'}V \\
  Z_{\mc{S}} @>>{\Theta}> Z_{\mc{S'}}
\end{CD}$$
Then
\begin{align}
  {\pi'}^W_*(\mc{V}_\mc{S'}) = \Theta_* \Big(\pi^W_*(\mc{V}_\mc{S})\Big), \label{eq:Onshell}
\end{align}
where $\mc{V}_S$ and $\mc{V}_{S'}$ are the integrands of $I_{\mc{S}}$ and $I_{\mc{S}'}$, respectively. In particular, if $Z_{\mc{S}} = Z_{\mc{S}'}$, then ${\pi'}^W_*(\mc{V}_\mc{S'}) = \pi^W_*(\mc{V}_\mc{S})$.
\end{Theorem}

\Proof This is an automatic consequence of the definitions, Lemma \ref{ThmWickCov}, and (\ref{Theta-pres}).\End

For $Z_\mc{S}$ compact, we can apply Definition \ref{DefMB} to obtain the Wick expansion of $I_{\mc{S}}$ about $Z_{\mc{S}}$:
\begin{align}
  W_{Z_{\mc{S}}}(\hbar) = \int_{Z_{\mc{S}}} \pi^W_*(\mc{V}_S) \label{def:WEgf} 
\end{align}
That is, the coefficients of $\pi^W_*(\mc{V}_S)$ are top-degree forms on $Z_{\mc{S}}$, and we integrate along $Z_{\mc{S}}$ to obtain a formal series in $\hbar$.

\begin{Theorem}
  The Wick expansion $W_{Z_{\mc{S}}}(\hbar)$ is independent of the choice of gauge-fixing slice $\mc{S}$.
\end{Theorem}

\Proof Equation (\ref{eq:Onshell}) shows that the integral (\ref{def:WEgf}) is independent of the choice of $\mc{S}$.\End

We may thus refer to $W_{Z_{\mc{S}}}$ as the \textit{gauge-fixed Wick expansion of $I$ about $Z$} with respect to $\mc{S}$. It is a well-defined formal series in $\hbar$, which requires a choice of gauge-fixing condition $\mc{S}$ for its construction but is independent of the choice made. It is built out of Wick contractions applied to the gauge-fixed integrand $\mc{V}_S$, and moreover its gauge-invariance arises from the covariance properties of Wick fiber integration. This involves formal algebra along the fibers rather than integration. Hence, the gauge-fixed Wick expansion serves as the correct finite-dimensional analog of gauge-fixed perturbative path integrals, as we will discuss in Section 4.

Having explained the Wick version of the slice Faddeev-Popov procedure, we now turn to the Wick version of the weighted Faddeev-Popov prcoedure, whose ansatz turns out to be more useful for the perturbative quantization of gauge theories. Here, in augmenting the integrand of $I$, we pick a gauge-fixing function $F: M \to \mc{Q}$ and weight function $\vp: \mc{Q} \to \R$ as before, with $\mc{S} = F^{-1}(q_0)$. We also make the additional assumption that $\vp$ is of the form $c_\hbar^{-1} e^{-h(q)/\hbar}$ where $h(x)$ has a nondegenerate critical point at $q_0$ and $c_\hbar$ is defined to be the Wick expansion of $\int dq e^{-h(q)/\hbar}$ (thus $c_\hbar^{-1}$ is a Laurent series in $\hbar^{1/2}$). 

We thus consider the weighted integral
$$I_{\vp, F} = \int dV \vp(F(x))\det (dF \circ \iota_x) f(x)e^{-S(x)/\hbar}.$$
It has Morse-Bott nondegenerate set $\mc{Z}_\mc{S}$ since $\vp(F(x))$ modifies the action $S(x)$ to a new one that is nondegenerate in the $G$-directions. 

\begin{Theorem}\label{ThmOnshellw}
 The Wick expansions of $I_{\vp,F}$ and $I_{\mc{S}}$ about $Z_{\mc{S}}$ agree. In
 fact, if $\tilde\pi$ is a fiber bundle structure for $Z_{\mc{S}}$ inside of $M$ (as in Definition \ref{Def:WFI}), then
 \begin{equation}
(\tilde\pi^W)_*\left(dV \vp(F(x))\det (dF \circ \iota_x) f(x)e^{-S(x)/\hbar}\right) \label{eq:WickFw}  
 \end{equation}
 is independent of the choice of $F$ and $\vp$ satisfying the requisite hypotheses. 
 Consequently, the Wick expansion of $I_{F,\vp}$ is also independent of $F$ and $\vp$.
\end{Theorem}

\Proof We know that the Wick expansion is independent of the choice of coordinates. So pull back the integrand of $I_{\vp,F}$ by $\rho: G \times \mc{S} \to M$. Since $h(F(x))$ is nondegenerate along the $G$-directions and constant along $\mc{S}$, the Wick expansion in the $G \times \mc{S}$ coordinate system factorizes in the $G$ and $\mc{S}$ directions. When we perform a Wick expansion in the $G$-direction, we have to Wick expand 
$$c_\hbar^{-1}\int_{G\cdot w} dV_G\, e^{-h(F(g\cdot w))/\hbar}\det (dF \circ \iota_{g\cdot w})$$
about $w \in \mc{S}$. By coordinate-independence ($g \mapsto F(g\cdot w)$ is a diffeomorphism in a neighborhood of the identity element of $G$) this equals the Wick expansion of 
$$c_\hbar^{-1}\int_{\mc{Q}} dq e^{-h(q)/\hbar}$$
which is $1$. The leftover Wick expansion involves $I_{\mc{S}}$.\End

Thus, the Wick version of the weighted Faddeev-Popov procedure only sees a ``local'' degree (equal to one) instead of the global degree that occurs in the integration version.

\section{The Wick Expansion and Integral Asymptotics}

In the previous sections, we considered the Wick expansion of an integral as a purely formal series in $\hbar$. This allowed us to consider properties of Wick expansions independently of the convergence properties of such integrals. In this section, we consider those integrals which are convergent for $\hbar$ along a ray in the complex plane. This makes
\begin{equation}
  I(\hbar) = \int d^dx f(x)e^{-S(x)/\hbar}, \label{Iasymp}
\end{equation}
a function of $\hbar$. We consider $\hbar$ along different rays because the cases $\hbar$ real and imaginary correspond to integrals arising from Euclidean and Lorentzian physics, respectively. Intermediate cases correspond to integrals obtained through Wick rotation. In what follows, we always assume $S(x)$ is real-valued.

As is well known, the Wick expansion provides an asymptotic expansion for $I(\hbar)$ as $\hbar \to 0$, in the sense of (\ref{asymp-series}), under suitable hypotheses. In this context, the method by which it is shown that the Wick expansion provides the correct asymptotics goes by several names, including saddle point approximation, stationary phase, or steepest descent. The asymptotic nature of the Wick expansion makes it unsurprising that the Wick expansion has the covariance properties that it has.  Indeed, the Wick expansion automatically inherits these properties from the corresponding ones for ordinary integrals when the former is an asymptotic expansion of the latter. However, what the previous sections show is that, essentially as a result of  Lemma \ref{LemmaIBP}, the covariance properties of the Wick expansion, with $\hbar$ a formal parameter, are of a purely algebraic nature and so hold without any additional hypotheses on the underlying integrand.


For completeness, we provide a proof of the asymptotic properties of the Wick expansion in the lemma below. We treat the case of nondegenerate critical points, everything being exactly analogous for the Morse-Bott case. First, we deal with the case when $f(x)$ has compact support and we work locally near a critical point $x_0$ of $S(x)$.  This is the most natural case, since for oscillatory integrals in which $\hbar$ is imaginary, the term $e^{-S(x)/\hbar}$ will not decay at infinity and thus one has to truncate the integral smoothly.

Write the Wick expansion as
$$W_{x_0}(\hbar) = e^{-S(x_0)/\hbar}\sum_{k =0}^\infty a_k\hbar^{k/2}.$$

\begin{Lemma}\label{LemmaAsymp}
Consider the integral
\begin{equation}
  I(\hbar) = \int d^dx\, f(x)e^{-\S(x)/\hbar}
\end{equation}
with $f(x)$ compactly supported. Suppose $S(x)$ has a unique nondegenerate critical point $x_0$ on the support of $f(x)$. Let $\hbar \to 0$ along a ray in the complex plane, where if $\mr{Re}\, \hbar > 0$, suppose further that $x_0$ is a minimum of $S(x)$. Then $W_{x_0}(\hbar)$ is an asymptotic series for $I(\hbar)$ in the sense that for every $N > 0$,
\begin{align}
I(\hbar) - e^{-S(x_0)/\hbar}\sum_{k < N} a_k\hbar^{k/2} = e^{-S(x_0)/\hbar}O(\hbar^{N/2}). \label{asymp-series}
\end{align}
\end{Lemma}

\Proof Without loss of generality, we can suppose $x_0 = 0$.  The asymptotics of $I(\hbar)$ are unchanged if we replace $I(\hbar)$ with
\begin{equation}
  I(\hbar) = \int d^dx\, \psi(x)f(x)e^{-\S(x)/\hbar}
\end{equation}
where $\psi(x)$ is any bump function that is identically one in some neighborhood of the origin. For $\mr{Re}\,\hbar > 0$, this is because for $x$ in the support of $f(x)$ and bounded away from $x_0$, $e^{-S(x)/\hbar} \leq e^{-S(x_0)/\hbar}e^{-C/\hbar}$ for some positive constant $C$. For $\mr{Re}\,\hbar = 0$, we use the standard integration by parts trick. Starting with $d = 1$, we have
\begin{align}
  \int dx\, [1-\psi(x)]f(x)e^{-\S(x)/\hbar} = \int dx\, [1- \psi(x)]f(x)\left(-\frac{\hbar}{S'(x)}\frac{d}{dx}\right)^N e^{-S(x)/\hbar} \label{ibp}
\end{align}
for arbitrary $N$. On the support of $[1-\psi(x)]f(x)$, we have that $S'(x)$ is nonzero, in which case integration by parts shows that the above integral is $O(\hbar^N)$. Since $N$ was arbitrary, such a term is asymptotically zero. For $d > 2$, one can adapt this technique to draw the same conclusion, see \cite[Chapter VIII]{Stein}.

The inverse function theorem allows us to choose local coordinates near the origin in which $\S(x) = \frac{1}{2}A(x,x)$. By making the support of $\psi(x)$ as small as we like, we can suppose that such coordinates have been choosen globally. This is a more convenient coordinate system to establish that the asymptotics of the function $I(\hbar)$ is equal to the Wick expansion of $I(\hbar)$ regarded as a formal integral. Since the Wick expansion is independent of the coordinate system chosen, our choice of coordinates does not affect the result.

Pick $\delta > 0$. Write
$$e^{\delta |x|^2}f(x) = P_N(x) + R_N(x)$$
where $P_N$ is the degree $N$ Taylor polynomial of $e^{\delta |x|^2}f(x)$ centered at the origin and $R_N$ the remainder. So then
\begin{align}
  \int d^dx\,\psi(x)f(x)e^{-S(x)/\hbar} &= \int d^dx\, P_N(x)e^{-\delta |x|^2}e^{-A(x,x)/2\hbar} \nonumber \\
  & \quad + \; \int d^dx\,[\psi(x)-1]P_N(x)e^{-\delta |x|^2}e^{-A(x,x)/2\hbar} \nonumber \\
  & \quad + \; \int d^dx\,\psi(x)R_N(x)e^{-\delta |x|^2}e^{-A(x,x)/2\hbar}. \label{3terms}
\end{align}
For the first term, we apply Wick's theorem using the quadratic form $2\delta|x|^2+A(x,x)/\hbar$, which has positive definite real part for $\delta$ small.  Sending $\delta \to 0$, we obtain leading terms of the Wick expansion. For the second term, if $\mr{Re}\,\hbar > 0$, we have pointwise exponential decay of the integrand. If $\mr{Re}\,\hbar = 0$, the same estimate used to control (\ref{ibp}) shows that the integral is $O(\hbar^M)$ for all $M > 0$. Finally, the third term of (\ref{3terms}) is $O(\hbar^{(d+N+1)/2})$. For $\mr{Re}\,\hbar > 0$, this follows from the rescaling $x \mapsto (\mr{Re}\,\hbar)^{1/2} x$ and using $R_N(x) = x^{N+1}\eta(x)$ with $\eta$ smooth. For $\mr{Re}\,\hbar = 0$, one has to work harder, see \cite[p. 335]{Stein}. Since $N$ was arbitrary, this establishes the lemma. \End

Next, we want to remove the hypothesis that $f(x)$ is compactly supported. Additionally, we want to impose conditions such that the sum of Wick expansion of $I(\hbar)$ at all of its critical points yields the full asymptotics of $I(\hbar)$. For simplicity, we assume $\S(x)$ has finitely many critical points $x_c$ and that all of them are nondegenerate. (The asymptotics of $I(\hbar)$ about degenerate critical points can be studied, but they cannot be analyzed via the Wick expansion.) Moreover, we assume $\S(x) \to \infty$ sufficiently rapidly at infinity. Thus, for $\mr{Re}\,\hbar > 0$, as long as $f(x)$ does not grow too quickly, (\ref{Iasymp}) is well-defined. For $\mr{Re}\,\hbar = 0$, the function $e^{-S(x)/\hbar}$ is no longer damping at infinity but highly oscillatory. Nevertheless, so long as $S(x)$ is suitably well-behaved (depending on $f(x)$), we can make sense of the regulated integral
\begin{align}
  I(\hbar) &= \lim_{\eps \to 0}I_\eps(\hbar) \nonumber \\
  &=: \lim_{\eps \to 0}\int d^dx\, \psi(\eps x)f(x)e^{-\S(x)/\hbar}. \label{osc}
\end{align}
Namely, we need $\frac{1}{S'(x)}$ and all its derivatives to decay sufficiently rapidly at infinity relative to $f(x)$, so that we can use integration by parts to control (\ref{osc}) uniformly in $\eps$.

\begin{Theorem}\label{ThmAsymp} Assume the above hypotheses and those of Lemma \ref{LemmaAsymp}. Let $\{x_c\}$ denote the set of critical points of $S(x)$ if $\mr{Re}\,\hbar = 0$ or else the set of minima of $S(x)$ if $\mr{Re}\,\hbar > 0$. Assume all the $x_c$ are nondegenerate. Then $\sum_{x_c}W_{x_c}(\hbar)$ is an asymptotic series for $I(\hbar)$.
\end{Theorem}

\Proof The assumed decay properties allow us to control the final two terms of (\ref{3terms}) as before, where we replace the function $\psi$ with $\psi(\eps x)$ and let $\eps \to 0$. These terms are $O(\hbar^{(d+N+1)/2})$ uniformly as $\eps \to 0$.$\;\square$

\section{Remarks on Quantum Field Theory}

We now apply our previous mathematical analysis to the setting of quantum field theory. We assume the reader has experience with quantum field theory so that we can be allowed to provide succinct commentary rather than a self-contained exposition. First, we review how the infinite-dimensional Wick expansion works in the setting of quantum field theory. After this, we carry out our main goal of explaining how a proper distinction between the Wick expansion and integration (as inspired by our previous finite-dimensional analysis) provides a conceptually sound treatment of formal path integral manipulations.

In quantum field theory, one wishes to apply the Wick expansion to a path integral, i.e., an integral over the space of field configurations\footnote{Properly speaking, this terminology is appropriate for Lorenztian theories in which there is a notion of time, hence the notion of a path. We use this terminology to include Euclidean theories.}. The path integral is formed out of a classical action $S$, from which we obtain the classical equations of motion by considering those configurations for which the action is stationary. One may then attempt to Wick expand the path integral about these stationary configurations by a formal application of Definition \ref{DefWE}\footnote{One almost always has a moduli of solutions to the classical equation of motion, whose finite-dimensional analog yields the Morse-Bott situation of Section \ref{SecMB}. We apply Definition \ref{DefWE} to compute Feynman diagrams evaluated on \textit{on-shell configurations}, i.e. those satisfying the equations of motion, which is analogous to performing Wick fiber integration $\pi^W_*$ (Feynman diagrams) and then considering $Z$ (on-shell configurations).}. Aside from an overall normalization, this is possible as long as we can make sense of the terms appearing in the Wick formula (\ref{WO}). The problem of course is that now the right-hand side of (\ref{WO}) is an integral of a product of distributions, since the inverse matrix $A^{ij}$ is to be replaced with an appropriate Green's function (the propagator) and the sum over indices become operator multiplication and integration. These are the familiar divergences appearing in Feynman diagrams. Thus, making sense of the Wick expansion, and indeed all other quantities in quantum field theory, require the proper use of \textit{regularization} and \textit{renormalization}.

There are three common methods one can use to regulate path integrals:
\begin{itemize}
  \item lattice regularization: replace the continuum theory with one defined on a (finite) lattice, so that the path integral becomes a product of ordinary integrals;
  \item propagator regularization: replace the integral kernel of the propagator with a smoothed out version, using e.g. a momentum cutoff or a heat kernel regulator;
  \item dimensional regularization: do not modify the propagator, but analytically continue the linear operator $\int d^dx$ for integer $d$ to complex $d$ (a rigorous treatment can be found in \cite{Col}\footnote{Note that in \cite{Col}, the dimensionally regularized integration operator $\int d^dx$ is only a linear operator when restricted to a suitable space of functions, e.g. rational functions. This is sufficient for the kinds of integrals that appear in translation-invariant field theories, whose integrands have Fourier transform equal to rational functions of momentum variables.}).
\end{itemize}
In all of these regularization schemes, there is a regulatory parameter $\eps$ that serves as an ultraviolet regulator which is removed as $\eps \to 0$. (One may also need an infrared regulator, which can always be fixed as one removes the ultraviolet cutoff.) For instance, $\eps$ can serve as a lattice spacing, the inverse of the momentum cutoff, or else one might work in dimension $d - \eps$. With a regularization scheme in place, this renders all integrals occurring in the regulated Wick expansion $W_\eps(\hbar)$ finite for $\eps > 0$. Thus, $W_\eps(\hbar)$ becomes a well-defined formal series in $\hbar$ since the Wick formula (\ref{WO}) generates well-defined coefficients at every order in $\hbar$, i.e. every loop order (modulo an overall normalization constant arising from a determinant, which we ignore from now on).

Next, renormalization begins by making the action depend on $\eps$, in which we write $S_\eps = S + CT_\eps$, a sum of the original action plus counterterms $CT_\eps$. Such counterterms, which diverge as $\eps \to 0$, are arranged so that the renormalized Wick expansion (the Wick expansion associated to $S_\eps$) has an $\eps \to 0$ limit as a formal power series in $\hbar$.

This is the general schema by which the standard path integral approach to perturbative quantum field theory proceeds. Each theory requires the appropriate technical execution of the above procedures, along with any additional requirements imposed by other considerations such as symmetry. The successful execution of such steps provides a perturbative \textit{definition} of a path integral, one which a priori is disconnected from trying to perform integration. From this, the basic principle behind making sense of path integral manipulations, in direct analogy of the Formal Integration Principle of Section 1, can be summarized as follows:\\

\noindent \textbf{Formal Path Integration Principle: }\textit{Any formal manipulation of a perturbatively defined path integral yields provisional identities whose legitimacy depend upon an analysis of the regularization and renormalization scheme employed.\\}

\noindent While those knowledgeable in quantum field theory are well-aware of this fact, it is unfortunate that this tacit philosophy is often obscured by formal notation or else interwoven with other procedures that have no rigorous analog. The approach we adopt in this paper yields a most natural elucidation of the Formal Path Integration Principle, since it cleanly separates the distinction between integration and formal integration already present in finite dimensions from the technicalities of regularization and renormalization occurring in infinite dimensions. 

We now investigate how a variety of path integral manipulations can be interpreted rigorously by conscientiously using the Wick expansion together with the above Formal Path Integration Principle. We proceed as follows. Section \ref{Sec:change} provides a warmup that shows how the different regularization schemes we outlined above work in generating the Wick expansion. Section \ref{Sec:Ward} shows how the subtleties involved in understanding Ward identities can already be gleaned from the finite-dimensional distinction between integration and the Wick expansion. In Sections \ref{SecGF} and \ref{SecEOM}, the theorems we proved in Section \ref{Sec:WEGF} are brought to their full light in terms of interpreting perturbative path integrals through the Wick expansion and not integration, and we explicitly mention some of the shortcomings in the literature.

\subsection{Change of variables in the path integral}\label{Sec:change}

For simplicity, we illustrate how to make sense of changes of variables in the path integral for the case of scalar field theories on $\R^d$, though everything we discuss readily carries over to more complicated theories. That is, we consider path integrals of the form
\begin{align}
  I = \frac{1}{Z}\int D\phi\,O(\phi) e^{-S(\phi)/\hbar} \label{pathintegral}
\end{align}
where
$$S(\phi) = \int d^dx \left[\frac{1}{2}\phi(x)\Delta\phi(x) + \frac{1}{2}m^2\phi(x)^2 + V\big(\phi(x)\big)\right],$$
$V$ and $O$ are some polydifferential functions of $\phi$, and $Z$ is a normalization constant. What follows is a toy computation to make explicit how regularization and the Wick expansion of path integrals intertwine. We then make comments about the more general situation afterwards.

To begin, consider a free massless theory with action
\begin{align*}
  S(\phi) = \frac{1}{2}\int d^dx\,  \phi(x)\Delta\phi(x).
\end{align*}
With $O \equiv 1$, the Wick expansion of (\ref{pathintegral}) is trivial (i.e. is a constant which we normalize to one) since there are no interaction terms. However, suppose we make a change of variables $\phi(x) \mapsto \phi(x) + \phi^3(x)$. The new action will no longer be a free theory and so will generate Feynman diagrams via a Wick expansion of the path integral about the zero field.

We can gain insight by seeing what happens in the one-dimensional case. The substitution $x \mapsto x + x^3$ does the following:
$$\frac{1}{\sqrt{2\pi\hbar}}\int dx e^{-x^2/2\hbar} \longrightarrow \frac{1}{\sqrt{2\pi\hbar}}\int dx(1 + 3x^2)e^{-\left(\frac{1}{2}x^2 + x^4 + \frac{1}{2}x^6\right)/\hbar}.$$
By Theorem \ref{LemmaCoord}, the Wick expansions of the two integrals about $x_0=0$ are identically the same (and in this case, the integrals are themselves convergent). But while the Wick expansion of the left-hand side is trivial, the right hand-side involves cancellations of diagrams to all orders in $\hbar$, since we have interaction terms. For instance, at first order in $\hbar$, the Wick expansion receives a contribution from $3x^2$ and $-\frac{x^4}{\hbar}$. This yields $3\hbar$ and $-3\hbar$, respectively, which cancel. Likewise, at order $\hbar^2$, we have to apply Wick's theorem to $-3x^2 \cdot \frac{x^4}{\hbar} + \frac{1}{2}\left(\frac{x^4}{\hbar}\right)^2 -\frac{x^6}{2\hbar}$, which yields $\hbar^2(-3\cdot 5!! + \frac{1}{2}7!! - \frac{1}{2}5!!) = 0$.

In quantum field theory, the combinatorial factors occurring in the Wick expansion are the same as those that appear in the Wick formula (\ref{WO}). It is the Wick contractions appearing in (\ref{WO}) that are affected by which regulator we choose. We illustrate what happens using all three methods previously discussed.

For lattice regularization, we have lattice points $x_i$ belonging to a finite-size box embedded within a rectilinear lattice with lattice spacing $\delta x$. Let $\Delta$ denote the lattice Laplacian with periodic boundary conditions (so that $\Delta = \Delta_{ij}$ is a matrix) and let $(\cdot,\cdot)$ denote the lattice inner product defined by
\begin{align*}
  (f,g) = \sum_i (\delta x)^df(x_i)g(x_i).
\end{align*}
The lattice partition function is then
\begin{align*}
Z = \int \prod_i d\phi(x_i)e^{-\frac{1}{2\hbar}(\phi, \Delta \phi)}.
\end{align*}
where $d\phi(x_i)$ is Lebesgue measure for the real variable $\phi(x_i)$. Under the change of coordinates $\phi(x_i) \mapsto \phi(x_i) + \phi^3(x_i)$, the integral changes to
\begin{align*}
\int \prod_i d\phi(x_i)\Big(1 + 3\phi^2(x_i)\Big)e^{-\left[\frac{1}{2}(\phi, \Delta \phi)+ (\phi, \Delta\phi^3) + \frac{1}{2}(\phi^3,\Delta\phi^3)\right]/\hbar}.
\end{align*}
Let $G_{ij}$ be the inverse matrix of $\Delta_{ij}$ restricted to the orthogonal complement of constant functions. In the continuum limit $\delta x \to 0$, we have
\begin{align*}
\sum (\delta x)^df(x_i) \to \int d^dx\,f(x),
\end{align*}
so that $G_{ij}(\delta x)^{-d}$ is the lattice version of the continuum Green's function:
\begin{align*}
\sum_j G_{ij}f(x_j) = \sum_j (\delta x)^dG_{ij}(\delta x)^{-d}f(x_j) \to \int d^dx\,G(x',x)f(x).
\end{align*}
Moreover, $G_{ij}(\delta x)^{-d}$ is the matrix we use to perform Wick contractions when we Wick expand.

At order $\hbar$, we have to consider Wick contractions of $\sum_i 3\phi^2(x_i)$ and the quartic interaction.
Wick contraction of  $\sum_i 3\phi^2(x_i)$ yields $3\sum_i G_{ii}(\delta x)^{-d}$. It is a regulated version of its divergent counterpart in the continuum limit, which is given by
\begin{equation}
  3(\delta x)^{-d}\sum_i (\delta x)^dG_{ii}(\delta x)^{-d}  \longrightarrow 3\delta^d(0) \int_M d^dxG(x,x) \label{eq:diag1}
\end{equation}
where the latter integral is performed over the torus $M$ given by the periodic identification of the box defining our lattice. Wick contraction of the quartic interaction yields
\begin{align}
  -\sum_i (\delta x)^d\phi^3(x_i)\Delta\phi(x_i) & \longrightarrow -3\sum_{i,j} (\delta x)^{d}\,G_{ii}(\delta x)^{-d}\Delta_{ij}G_{ji}(\delta x)^{-d} \label{eq:diag2}\\
  &= -3\sum_i G_{ii}(\delta x)^{-d}.
\end{align}
This exactly cancels out the previous diagram.

For propagator and dimensional regularization, one starts with the continuum theory. The path integral transforms as
$$\int D\phi \left(1 + \delta^{(d)}(0)\int d^dx\,3\phi^2(x)\right)e^{-\left[\frac{1}{2}(\phi, \Delta \phi)+ (\phi, \Delta\phi^3) + \frac{1}{2}(\phi^3,\Delta\phi^3)\right]/\hbar}.$$
One can justify this formal manipulation on the basis of the formal continuum limit of the lattice theory above. The difference is that previously, all quantities were well-defined on a finite lattice whereas now one has to ``undo" these formal manipulations with a regularization scheme different from the lattice.

For propagator regularization, this means regulating
\begin{align}
  G(x,y) &= \frac{1}{(2\pi)^d}\int d^dp\, e^{i(x-y)}\frac{1}{|p|^2} \label{eq:FT1}\\
  \delta^{(d)}(0) &= \frac{1}{(2\pi)^d}\int d^4 p \label{eq:FT2}
\end{align}
by taming the integrands in these expressions. In the present situation, the two Feynman integrals we have to consider are $3\delta^{(d)}(0)\int d^dx\,G(x,x)$ and $-3\int d^dx\,G(x,x)\Delta G(x,x)$ arising from Wick contracting $\delta^{(d)}(0)\int d^dx\,3\phi(x)^2$ and $-\int d^dx\,\phi\Delta\phi^3/2\hbar$, respectively. Formally,
\begin{equation}
  \delta^{(d)}(0) - \Delta G(x,x) = 0 \label{eq:delta-G}
\end{equation}
and so the two Feynman integrals cancel. To make (\ref{eq:delta-G}) more meaningful, we can regulate the theory in such a way that if we replace $\delta^{(d)}(0)$ and $G$ in (\ref{eq:delta-G}) with regulated versions, call them $\delta^{(d)}_\eps$ and $G_\eps$, we still get exact cancellation. More generally, what we need is that
\begin{equation}
  \lim_{\eps \to 0}\Big(\delta_\eps^{(d)}(0) - \Delta G_\eps(x,x)\Big) = 0. \label{eq:delta-Ge}
\end{equation}
Such a regularization procedure makes rigorous the cancellation of two infinite quantities. It is straightforward to obtain (\ref{eq:delta-Ge}) by inserting the appropriate momentum cutoff in the integrands of (\ref{eq:FT1}) and (\ref{eq:FT2}).

In dimensional regularization, one obtains a Wick expansion using the integration operator $\int d^dx$ with $d$ complex. In this setting, it turns out there are no power divergences, i.e. $\int d^dp\, |p|^{2\alpha} = 0$ for all real $\alpha$. Thus, both $\delta^{(d)}(0)$ and $\Delta G(x,x)$ are zero individually, i.e. both diagrams under consideration vanish. Strange as this may seem, so long as $\int d^dx$ is a consistent\footnote{See also \cite{CL} for some difficulties with dimensional regularization if not done properly.} linear operation, this provides one with a well-defined \textit{rule} for performing the Wick expansion. While the author has personal reservations about using dimensional regularization due to this strange elimination of divergences, it is the presence of such fortuitious eliminations of divergences which makes dimensional regularization popular\footnote{In addition, dimensional regularization preserves gauge symmetry, which makes it a standard choice for quantizing gauge theories.  We should note that dimensional regularization, at least as presented here, only applies to translation-invariant theories and thus has rather limited applicability. In contrast, lattice regularization and propagator regulation via heat kernel methods are robust and can be applied to general situations. For some comparisons between different regularization schemes in specific models, see e.g. \cite{LNP, BLZ}.}.\\

Having discussed the above example, it is no more difficult conceptually to consider transformations of the form $\phi \mapsto \phi + F(\phi)$ with $F$ a polynomial that is at least quadratic. Such a transformation is invertible as a power series in $\phi$. One can inspect the diagrams one generates when performing such a change of variables and inspect how regularization meshes with the terms of the Wick expansion. Such an inspection has to be done on a case by case basis, with formal manipulations of the path integral being a way of delaying, or in many cases, masking this inspection. One could also consider transformations of $\phi$ in which the linear term is not the identity, but some general linear transformation $L\phi$. In the Wick expansion, one picks up a determinant factor from this linear transformation. This must also be regulated appropriately.

\subsection{Ward identities and Schwinger-Dyson equations}
\label{Sec:Ward}

One often wants to show that the path integral of a Lie derivative is zero, where the Lie derivative arises from some infinitesimal symmetry or infinitesimal change of variables. The resulting identities are referred to by various names depending on their context, which we recall in Table 1. Despite the many names, it is customary to refer to the general collection of identities obtained through infinitesimal variations of the path integral as simply Ward identities.

\setlength{\extrarowheight}{5pt}
\begin{table}[h]
\begin{tabularx}{\textwidth}{|l|X|}\hline
Identity & Operation  \\ \hline
  Schwinger-Dyson equations & varying a scalar field by translation $\phi(x) \mapsto \phi(x) + \eps(x)$ \\
  Ward identity & varying the electron by its phase $\Psi \to e^{i\eps(x)}\Psi$ in quantum electrodynamics \\
  Slavnov-Taylor identities & varying all fields by the BRST operator in gauge-theories \\ \hline
\end{tabularx}\vspace{.15in} \caption{Various identities occurring in quantum field theory.}
\end{table}

For the case of Schwinger-Dyson equations, the identity we obtain is usually expressed in the form
\begin{equation}
  \left<\frac{\delta O(\phi)}{\delta \phi}\right> = \frac{1}{\hbar}\left<O(\phi)\frac{\delta S(\phi)}{\delta \phi}\right>, \label{SD}
\end{equation}
for $O$ a general observable, where $\left<O\right>$ is defined by (\ref{pathintegral}). This equation is interpreted as a ``quantum equation of motion", since for example taking $O \equiv 1$, this equation says that the expectation of $\frac{\delta S(\phi)}{\delta \phi}$ is zero. Equation (\ref{SD}) is obtained by declaring (i) the path integral is invariant under the translation change of variables $\phi(x) \mapsto \phi(x) + \eps(x)$; (ii) the ``measure" $D\phi$ is invariant under translation. Differentiating with respect to $\eps$ at $\eps = 0$ yields
\begin{align}
  0 &= \delta I \\
  &= \frac{1}{Z}\int \delta\left(D\phi\,O(\phi) e^{-S(\phi)/\hbar}\right) \label{eq:intexact=0}\\
  &= \frac{1}{Z}\int D\phi \left(\frac{\delta O(\phi)}{\delta\phi} - \hbar^{-1}O(\phi)\frac{\delta S(\phi)}{\delta\phi}\right)e^{-S(\phi)/\hbar}.
\end{align}

We see that (\ref{SD}) is a formal analogue of (\ref{eq:deriv}). As such, (\ref{SD}) has to be interpreted in the sense of formal power series in $\hbar$, in which each side generates its own Wick expansion. Thus, equation (\ref{SD}), which holds formally, holds legitimately if one can arrange for regularization and renormalization (as done in the previous section) in the Wick expansion to be carried out so as to render (\ref{SD}) true. For instance, using lattice regularization, (i) and (ii) are obviously true since Lebesgue measure at each lattice site is translation invariant. In fact, there will in general be correction terms to (\ref{SD}) due to the fact that nonlinear functionals of the field need to be renormalized.

It is important to note that many treatments drop the explicit dependence on $\hbar$ in (\ref{SD}), treating $\hbar$ as a number (such as $i$ \cite[Ch 9.6]{PS}). This blurs the formal power series nature of the involved quantities. Moreover, we also have to remember that the Wick expansion is usually about a moduli of configurations. In the finite-dimensional situation of Section 2, the analogue of Lemma \ref{LemmaIBP} would hold only if one could determine that the integral of a total derivative on the on-shell space $Z$ vanishes as well. This is also blurred in the formal notation (\ref{SD}). Any rigorous interpretation of the provisional equation (\ref{SD}) must take these considerations into account\footnote{There are algebraic approaches to perturbative quantum field theory with a cohomological emphasis, in which one considers algebras of observables equipped with a differential (see e.g. \cite{CG, FR}). With the appropriate setup in place, one regards a quantum expectation as taking the cohomology of this differential. In the finite-dimensional setting, this would be analogous to replacing integration over the on-shell space $Z$ in Definition \ref{DefMB} with passing to cohomology. This automatically makes integration by parts valid, i.e., the expectation of an exact term is zero, without any hypotheses on $Z$. However, such a cohomological setup has the disadvantage of losing connection with integration, even in the rigorous finite-dimensional situation.}.

The other Ward identities can be handled in a conceptually similar manner\footnote{For those theories involving ghosts and odd symmetries, to accurately mimic those theories in the finite-dimensional setting, one would have to invoke a version of the Wick expansion which includes fermionic variables. However, as fermionic integration is already purely algebraic, one can readily extend our finite-dimensional analysis of the bosonic Wick expansion to the case of supermanifolds by adapting the differento-geometric framework of supermanifolds in \cite{DM}.}. The fact that one varies by a symmetry of the action means that one expects to obtain identities without terms proportional to the equation of motion. However, if the ``measure" $D\chi$ on the space of fields $\chi$ is not invariant with respect to the symmetry, one obtains an \textit{anomaly} which is a correction term appearing as a Lie derivative of the measure. In the finite-dimensional Wick expansion, such a Lie derivative can be computed using the algebraic rule $\L_V = d\iota_V$, in which one obtains a divergence factor $\pd_i V^i(x)$ associated to $V = V^i(x)\pd_x^i$. Indeed,
$$\L_X(dx^1\cdots dx^d) = \pd_i V^i(x) dx^1 \cdots dx^d.$$
It is the appearance of this divergence factor (suitably regulated) that can be generalized to the infinite-dimensional setting without  requiring the existence of a measure for integration.

The well-known chiral anomaly of QED falls under this analysis, in which an anomaly term arises from the local index density of a chiral Dirac operator \cite[Ch 19.2]{PS}. A general framework of how to properly treat the change in the ``measure" $D\chi$ under a symmetry in continuum perturbative quantum field theory is provided by the Batalin-Vilkovisky formalism as developed in \cite{Cos}. For a concrete example of this formalism in the context of nonlinear sigma models, see \cite{N}.

Altogether, we see how various subtleties involved with obtaining and interpreting Ward identities can be inferred from the algebraic features of the Wick expansion. Trying to discuss these issues in terms of integration causes difficulties when trying to justify the integration of an total derivative being zero in (\ref{eq:intexact=0}) or the existence of a measure that yields an anomaly.

\subsection{Gauge-fixing of Feynman amplitudes and path integrals}\label{SecGF}

As an application of our analysis of the Wick expansion in Section 2, let us provide some insights into the gauge-fixing procedures done in quantum field theory. 

Our first goal is to provide a finite-dimensional interpretation of the gauge-invariance of on-shell Feynman amplitudes. Accomplishing this was in fact, one of our motivations for this paper, since conventional path integral methods obscure why the on-shell condition is crucial. The specific example we have in mind arises from quantum electrodynamics (QED). Here, there is a natural family of gauge-fixing conditions parametrized by $\xi \geq 0$ with corresponding propagator
\begin{equation}
  \frac{1}{p^2}\left(g^{\mu\nu} - \frac{(1-\xi)p^\mu p^\nu}{p^2}\right). \label{eq:xi-gauge}
\end{equation}
Inspection shows that it is the Wick version of the weighted Faddeev-Popov procedure (adapted to quantum field theory) that is used to obtain the propagator (\ref{eq:xi-gauge}). Indeed, inserting the weight $\exp\Big(-\int \frac{1}{2\xi\hbar}(\pd_\mu A^a_\mu)^2\Big)$ into the QED path integral introduces the $\xi$-term in (\ref{eq:xi-gauge}). Mimicking Theorem \ref{ThmOnshellw} in this infinite-dimensional context, the corresponding gauge-fixing functional is $F(A^a_\mu) = \pd_\mu A^a_\mu$ and varying $\xi$ corresponds to varying the weight $\vp(\cdot) = e^{-|\cdot|^2/2\xi\hbar}$. Theorem \ref{ThmOnshellw} shows that Wick fiber integration (generating Feynman diagrams) and then restriction to the gauge-fixed critical configurations $Z_{\mc{S}}$ of the action $S$ (on-shell evaluation)\footnote{Wick fiber integration yields a top-degree form on $Z_{\mc{S}}$, but we can identify it with a function on $Z_{\mc{S}}$ once we choose a fixed reference $G$-invariant volume on $Z$. The Taylor series of the resulting function yields polynomials whose evaluation on $Z_{\mc{S}}$ correspond to evaluating Feynman diagrams with elements of $Z_{\mc{S}}$ placed on external legs.} is independent of the variation of $\vp$. Of course, as with the previous section on changes of variables, one always has to inspect how regularization and renormalization affect the translation of any result about finite-dimensional Wick expansions to infinite dimensional ones. Since our goal here is to highlight the conceptual fitness of the Wick expansion, we leave the details to the industrious reader. 

By contrast, the conventional description of gauge-fixing which proceeds by way of analogy with the finite-dimensional integration version of the Faddeev-Popov procedure is misleading. Indeed, the only finite-dimensional gauge-invariant regularization of gauge theories (to the author's knowledge) involves working with group-valued instead of Lie-algebra valued fields on a lattice. Since in most instances, the continuum limit of the lattice formulation has not been rigorously shown to approach the continuum theory, it is unclear how the integration Faddeev-Popov procedure on the lattice is related to the Wick Faddeev-Popov procedure in the continuum \cite{Sha}. Next, the Feynman diagrams generated from a gauge-fixed action do not depend on topological features of the group of gauge transformations or of the gauge-fixing function $F$; the diagrams are generated purely algebraically from the gauge-fixed action. In finite dimensions, we saw that such a feature is true of the Wick version of the weighted Faddeev-Popov procedure but not of the integration version for which a topological degree term appears. Hence, once more it is the Wick expansion, not integration, that is most conceptually fit for describing how gauge-fixed path integrals are  perturbatively evaluated.

On the other hand, there are contexts in which the integration version of the Faddeev-Popov procedure is relevant to being able to evaluate path integrals nonperturbatively. Here,  we have in mind the computation of the path integral formed out of the Polyakov action in string theory (as performed in \cite{Jas}). Whereas the Wick version of the weighted procedure is used to yield well-defined Feynman rules in perturbation theory as above, the integration version of the slice procedure (Theorem \ref{ThmFP.R}) rewrites a path integral on a large space in terms of an integral on a much smaller space using quantities that one can regularize, namely the Faddeev-Popov determinant as defined by (\ref{detA}) or (\ref{det1/2}). In particular, if $M$ denotes the space of Riemannian metrics on a surface $\Sigma_h$ of genus $h$, and $G$ is the group of diffeomorphisms and conformal rescalings, then $M/G$ is the (finite-dimensional) moduli space of genus $h$ Riemann surfaces, and the slice Faddeev-Popov method converts the ill-defined path integral over $M$ to a sensible (nonperturbatively defined) integral over a slice $\mc{S}$. Here, we can see why the Riemannian version of the slice method (Theorem \ref{ThmFP.R}) is more useful than the implicit version (Theorem \ref{ThmFP.i}). The latter directly involves volume forms (e.g. Lebesgue measure), which do not always generalize to the infinite-dimensional setting. But for the former, since infinite-dimensional spaces can still be endowed with metrics, the corresponding Faddeev-Popov determinant can be assigned meaning via a suitable regularization, which allows for an evaluation of the resulting gauge-fixed path integral.

Altogether, we see how the distinct Wick and integration versions of the Faddeev-Popov procedure in finite-dimensions provide differing insights into how to proceed in quantum field theoretic contexts.

\subsection{Eliminating fields by their equation of motion}\label{SecEOM}

Often times, a field only appears quadratically in a path integral and one wishes to ``integrate out" this field. Two well-known instances of this are the Nakanishi-Lautrup field $h$ occurring in BRST gauge-fixing and the $B$-field in $BF$-theories. In the first instance, one wishes to perform the path integral
\begin{align}
  \int Dh\, e^{\int \left[(\pd_\mu A_\mu^a)h^a + \hbar\xi h^ah^a/2\right]} = e^{-(\pd_\mu A^a_\mu)^2/2\xi\hbar} \label{eq:nakanishi}
\end{align}
so as to insert the gauge-fixing term $e^{-(\pd_\mu A_\mu^a)^2/2\xi\hbar}$ in the path integral over $A^a_\mu$ (we ignore overall constants in front of path integrals in what follows). In the second instance, one wishes to convert Yang-Mills theory to a first order theory by writing
\begin{align}
  \int dA e^{-\frac{1}{2\hbar}\int F^a_{\mu\nu} \wedge * F^a_{\mu\nu} } = \int dAdB e^{\left(\frac{\hbar}{2}\int B^a_{\mu\nu} \wedge *B^a_{\mu\nu} + \int B^a_{\mu\nu} \wedge *F^a_{\mu\nu}\right)}.
\end{align}
The usual physics justification for these procedures is that one is either performing Gaussian integration or else eliminating a field using its equation of motion:
\begin{align*}
  h^a &\mapsto -\pd_\mu A_\mu^a/\xi\hbar\\
  B^a_{\mu\nu} &\mapsto -F^a_{\mu\nu}/\hbar.
\end{align*}
The latter justification is a notational shortcut for the former justification, which in turn is improper from the point of view of honest integration, since the integrals over $h^a$ and $B^a$ involve the exponential of a positive definite ($\xi > 0$) or else imaginary (in Lorentzian signature) quantity. Hence, the correct justification, comes from acknowledging that these procedures only involve the Wick expansion, which is well-defined regardless of the signature of the quadratic part of the action.

We can abstract the above analysis to the finite-dimensional case as follows. Let $T: X \to Y$ be an invertible linear map between inner product spaces of dimension $d$, and $V: X \to Y$ a polynomial map that is at least quadratic. We have the action
\begin{equation}
 S(x) = \frac{1}{2c}|Tx + V(x)|^2 \label{eq:SOaction}
\end{equation}
where $|\cdot|$  is the norm associated to the inner product $\left<\cdot,\cdot\right>$ on $Y$. One can Wick expand $\int d^dx\, e^{-S(x)/\hbar}$ about $x_0 = 0$. On the other hand, we can introduce an auxiliary variable $y \in Y$ and consider an equivalent ``first-order formulation" (we think of $T$ as being a first order operator).

Namely, consider the first order action
\begin{equation}
S_{FO}(x,y) = -\Big(\frac{c}{2}\left<y,y\right> + \left<Tx + V(x),y\right>\Big) \label{eq:FOaction} 
\end{equation}
The quadratic part of the action $-\frac{1}{2}\Big(c\left<y,y\right> + 2\left<Tx,y\right>\Big)$ can be identified with the matrix
\begin{align*}A =
  -\begin{pmatrix}
    0 & T^* \\
    T & c
  \end{pmatrix}
\end{align*}
via the inner product on $X$ and $Y$. Its inverse is given by
\begin{align*}
  A^{-1} = \begin{pmatrix}
    c(T^*T)^{-1} & -T^*(TT^*)^{-1} \\
    -T(T^*T)^{-1} & 0
  \end{pmatrix}.
\end{align*}
This yields a complicated set of Feynman rules if we Wick expand $S_{FO}$ about $(x,y)=0$. On the other hand, we know that the Wick expansion is invariant under changes of coordinates. Since
$$-\left(\frac{c}{{2}}\left<y,y\right> + \left<Tx,y\right>\right) = -\frac{c}{2}\left|y + Tx/c\right|^2 + \frac{1}{2c}|Tx|^2,$$
the change of variables $y \mapsto y - Tx/c$ diagonalizes the quadratic part of $S(x,y)$. Furthermore, being upper triangular, it is also a volume-preserving transformation. Thus the Wick expansion of $S_{FO}(x,y)$ is equivalent to the Wick expansion of
$$S(x,y) = -\frac{c}{2}|y|^2 - \left<V(x),y\right> + \frac{1}{2c}\Big(|Tx|^2 + 2\left<T(x),V(x)\right>\Big)$$
Since the quadratic part of $S(x,y)$ is diagonal, the Wick contractions of the $x$ and $y$ variables decouple. Wick expanding with respect to the $y$ variables first, we obtain
\begin{align}
  \int d^dx d^dy\, e^{-S(x,y)/\hbar} & \longrightarrow \left(-\frac{2\pi\hbar}{c}\right)^{d/2}\int d^dx\, e^{-|V(x)|^2/2c\hbar}e^{-\big(|Tx|^2 + 2\left<T(x),V(x)\right>\big)/2c\hbar} \label{eq:intouty1}\\
  &=  \left(-\frac{2\pi\hbar }{c}\right)^{d/2} \int d^dx\, e^{-S(x)/\hbar}. \label{eq:intouty2}
\end{align}
Thus, we recover the Wick expansion of $S(x)$ up to an overall normalization. (In the first line, we regard $\left<V(x),y\right>$ as an interaction. While this is not at least cubic in $y$, we can apply Wick's theorem and get a well-defined series expansion in the formal variable $\hbar$ and in the $x$-variable, since $V(x)$ is at least quadratic and so has no constant term.) Notice however, that (\ref{eq:intouty1})--(\ref{eq:intouty2}) can only be interpreted as a formal Wick expansion. If $\Re c < 0$, then (\ref{eq:intouty2}) has the wrong sign for doing a Gaussian integral; if $
\Re c > 0$, then the integration over $y$ in (\ref{eq:intouty1}) has the wrong sign. In other words, the sign of $c$ is always incompatible with either the $y$ or $x$ integral being convergent. It is unfortunate however that standard treatments of ``integrating out'' auxiliary fields describe this step (or some variation of it) as an honest integration procedure rather than as a Wick expansion. 

For a concrete instance, we can examine equation (15.7.5) from Weinberg's textbook \cite{Wein}, in which the gauge-fixing term $\exp(-\frac{i}{2\xi}\int (\pd_\mu A^a_\mu)^2)$ is described as arising from the ``Fourier integral''
\begin{equation}
\exp\left(-\frac{i}{2\xi}\int (\pd_\mu A^a_\mu)^2\right) = \int dh \left[\exp\left(\frac{i\xi}{2}\int h^ah^a\right)\exp\left(i\int \pd_\mu A^a_\mu h^a\right)\right] \label{eq:Fourier}.
\end{equation}
The above equation differs from (\ref{eq:nakanishi}) by factors of  $i$ due to working in Minkowski space instead of Euclidean space. One can experiment with how to interpret Weinberg's statement about (\ref{eq:Fourier}) being a ``Fourier integral''. Indeed, since $e^{\frac{i\xi x^2}{2}}$ is not a decaying Gaussian, one cannot literally perform the above integral, even its finite-dimensional analog. One could try to regulate (\ref{eq:Fourier}) by say endowing all occurences of $i$ in (\ref{eq:Fourier}) with a small negative real part, i.e., replace $i$ with $i - \eps$ (this regularization has the desirable property that the right-hand side of (\ref{eq:Fourier}) remains the exponential of a BRST exact term). This would make the right-hand side of (\ref{eq:Fourier}) damping but then the left-hand side acquires a small positive real part, and so the same convergence issue reappears if one is to take the interpretation of integration literally. The Wick expansion however is immune to such issues of convergence, and is thus the actual procedure upon which the above steps relie.\\

\addcontentsline{toc}{subsection}{4.5.\;\; Perturbative versus constructive QFT}


\noindent 4.5. \textbf{Perturbative versus constructive QFT.\,} There is of course much more to quantum field theory than the perturbative approach, and so we round out our discussion with some brief remarks to place our perturbative analysis in a more complete context.

To begin, we note that the value of the Wick expansion in the perturbative approach to a quantum field theory stems from the tacit assumption that the resulting formal series should provide some kind of approximation to the full theory, namely, an asymptotic series. As one can glean from the finite dimensional setting, this can only be true provided one can control the behavior of the path integral for large fields. Recall that Lemma \ref{LemmaAsymp} shows that the Wick expansion provides asymptotics for an integral localized via a smooth cutoff to a neighborhood of its critical points. Via Theorem \ref{ThmAsymp}, this boosts to an asymptotic expansion of the full integral given sufficient decay of the integrand at infinity. In infinite dimensions, establishing decay of the action in the path integral for large fields is of course much more nontrivial. This is quite evident in the case of nonabelian gauge theories, where gauge-invariance makes it difficult to establish bounds on the nonlinear terms of the Yang-Mills action when the gauge-field is large.

Such large field problems pose a obstacle in relating the Wick expansion of perturbatively defined path integrals to the full path integral, whatever that might be. But even if one ignores this issue and takes the Wick expansion as the starting out, one still has the problem of deciding what the Wick expansion is an asymptotic series of. In finite dimensions, $I(\hbar)$ can always be made a well-defined function of $\hbar$, either by taking $f(x)$ compactly supported or imposing the necessary decay properties. In perturbative quantum field theory, one does not start off with a well-defined integral but with the formal Wick expansion defined via the many intermediate steps of regularization, renormalization, and possibly other tricks. The choice of a function whose asymptotics are governed by such a formal series is not unique, since e.g. functions like $e^{-1/\hbar}$ are asymptotically zero.  Techniques for producing a canonical function whose asymptotics are given by a particular series (this provides a reverse to the diagonal arrow in Figure 1) include such methods as Borel summation. If one can successfully use such methods to sum a divergent asymptotic series to a function\footnote{This has been successfully carried out for scalar field theories in dimensions two and three \cite[Chapter 23.2]{GJ}.}, such a function could then be viewed as some truncated version of the path integral localized to fields near the Wick expansion locus. It would seem reasonable to conclude that such an analysis is necessary in order for one to connect purely formal perturbation theory with something that deserves to be called integration. See also \cite{DU} for some related  perspectives arising from ``resurgence theory'', in which it is hoped that the ambigiuities that arise from attempting to Borel sum a perturbative series encode essential features of nonperturbative physics.

We should of course note that in special cases, insight has provided \textit{suggestions} of how to \textit{define} the full path integral. We have in mind the remarkable insights of Witten, who (to name just two examples) used localization methods in symplectic geometry to offer an (alternative) construction of the Yang-Mills path integral in two dimensions \cite{Wit2D} and insights from topology to define the Chern-Simons path integral for compact, oriented $3$-manifolds \cite{WitCS}. These insights go well beyond what perturbation theory can do alone, in fact even bypassing what would otherwise be the analytic problem of defining measures on infinite dimensional spaces\footnote{Path integrals for $0+1$-dimensional theories are realized by the Wiener measure on the space of continuous paths. Likewise, in two dimensions one can construct path integral measures for general scalar field theories \cite{GJ}. Dimensions three and higher have proven to be extremely difficult.} or some modification thereof\footnote{For example, in two-dimensional Euclidean Yang-Mills, one can define the theory in terms the expectation values it assigns to Wilson loop observables, which one can interpret as a measure on the space of connections modulo gauge \cite{Levy}.}. One of the preeminent challenges to placing quantum field theory on firm mathematical foundations is to make sense of the path integral from first principles, thereby bridging the gap between formal perturbation theory, which is limited to the purely formal setting of series expansions, and the few specialized instances in which external insights provide a serendipitous definition of the path integral. There are of course formulations of quantum field theory that make no use of the path integral, but that is outside the scope of this article.

\section{Conclusion}

We provided a basic study of the Wick expansion in finite dimensions and discussed
its relation to integration. We then discussed how formal manipulations of perturbatively defined path integrals in quantum field theory are more properly understood through generalization of the finite-dimensional Wick expansion and not integration. This provides a rigorous treatment of such formal manipulations in a conceptually sound manner. We concluded with a discussion of constructive quantum field theory to show that there is still much to be desired in making mathematical sense of the much wider features of quantum field theory that lie outside of perturbation theory.

\end{document}